\documentclass[secnumarabic,
               amssymb, 
               nobibnotes, 
               reprint,
               pra]{revtex4-2} 
               
\usepackage{soul} 

\usepackage[T1]{fontenc}
\usepackage[utf8]{inputenc}
\usepackage{hyperref}  
\usepackage{multirow}  
\usepackage{subfigure} 
\usepackage{array}
\usepackage{booktabs} 
\usepackage{siunitx}

\usepackage{bbm}       
\usepackage{bm}        
\usepackage{mathtools} 
\usepackage{braket}    

\usepackage[T1]{fontenc}
\usepackage{xcolor} 
 
\definecolor{themeBlue}{RGB}{38,70,83}
\definecolor{themeGreen}{RGB}{42,157,143}
\definecolor{themeYellow}{RGB}{233,196,106}
\definecolor{themeOrange}{RGB}{244,162,97}
\definecolor{themeRed}{RGB}{221,101,81}

\hypersetup{
      breaklinks=true,  
      colorlinks=true,
      urlcolor=themeBlue,
      linkcolor=themeRed,
      citecolor=themeGreen,
} 

\setstcolor{themeRed}

\begin{document}

\title{Reduced density matrix approach to ultracold \\ few-fermion systems in one dimension}%

\author{Mitchell J. Knight}%
\email{knightm1@student.unimelb.edu.au}
\affiliation{School of Physics, University of Melbourne, Parkville, 3010, Australia}
\author{Harry M. Quiney}
\affiliation{School of Physics, University of Melbourne, Parkville, 3010, Australia}
\author{Andy M. Martin}
\affiliation{School of Physics, University of Melbourne, Parkville, 3010, Australia}
\date{Spetember 2021}%
\begin{abstract}
The variational determination of the two-fermion reduced density matrix is described for harmonically trapped, ultracold few-fermion systems in one dimension with equal spin populations. This is accomplished by formulating the problem as a semi-definite program, with the two-fermion reduced density matrix being subject to well-known $N$-representability conditions. The ground-state energies, as well as the density, pair-correlation function, and lower-order eigenvalues of the two-fermion reduced density matrix of various fermionic systems are found by utilising an augmented Lagrangian method for semi-definite programming. The ground-state energies are found to match well to those determined by full-configuration interaction and coupled-cluster calculations and the density, pair-correlation function, and eigenvalue results demonstrate that the salient features of these systems are well-described by this method. These results collectively demonstrate the utility of the reduced density matrix method firstly in describing strong correlation arising from short-range interactions, suggesting that the well-known $N$-representability conditions are sufficient to model ultracold fermionic systems, and secondly in illustrating the prospect of treating larger systems currently out of the reach of established methods.
\end{abstract}
\maketitle

\section{Introduction}

The study of few-fermion systems in quantum mechanics began in earnest immediately after the discovery of the Schr\"odinger equation in 1925 \cite{Schrodinger1926}. The first systems to be analysed were simple atoms and molecules. The treatments were extended to larger systems as the computational power increased and the theoretical methods became more well-founded. In more recent years, experimental data have permitted the study of more exotic few-fermion systems such as quantum dots \cite{Kouwenhoven2001,Zumbuhl2004,Fasth2007,Hanson2007} and trapped, highly-correlated ultracold fermionic gases \cite{Tonks1936,Girardeau1960,Busch1998,Blume2007,vonStecher2008,Blume2009,Rubeni2012,Sowinksi2013,Grining2015,Pecak2017,Sowinski2019,Bloch2008,Giorgini2008,Guan2013,Kinoshita2004,Haller2009,Guan2013,Zurn2012,Wenz2013,Murmann2015,Zurn2013}. The ongoing theoretical efforts to accurately and efficiently capture the ground-state properties and dynamics of these systems coupled with ever-growing experimental capacities has ensured that the field of few-fermion systems retains immense interest. 

A main goal of many-body quantum mechanics is the accurate description of correlation in a system, which constitutes the intricate interactions that are not captured by a mean-field approach. In typical atomic and molecular systems, long-range Coulombic interactions are responsible for correlation in the system. However, in ultracold systems, short-range, point-like interactions dominate and it is these characteristic interactions that give rise to the exotic effects exhibited by these systems. An apt many-body scheme would naturally capture the effects of correlation in a system regardless of the nature of the interaction. However, there is no \textit{a priori} reason that a scheme which accurately captures correlation from long-range effects will also capture correlation from short-range effects to the same degree of accuracy.

Fortuitously, many schemes that arose out of a need to capture correlation arising from Coulombic interactions are equally applicable to short-range interactions. Many of these schemes are known as \textit{post}-Hartree-Fock (HF) methods as they utilise the non-correlated HF ground-state as an initial step to construct a wavefunction solution to the Schr\"odinger equation. For example, the coupled-cluster (CC) method has seen wide success in atomic and molecular calculations \cite{Coester1958,Cizek1966,Cizek1971,Paldus1972,Bartlett1981,Bartlett1989,Bartlett2007} and, more recently, has been shown to accurately capture the ground-state of ultracold bosonic \cite{Cederbaum2006,Alon2006} and fermionc systems \cite{Grining2015,Grining2015a}, thus demonstrating the practical generality of the method. 

Included in the myriad techniques that constitute \textit{post}-HF methods is the technique of full-configuration interaction (FCI) which variationally finds an exact solution of the Schr\"odinger equation in the Hilbert space spanned by the finite basis set used. This method is equivalent to a direct diagonalisation of the Hamiltonian and serves as a benchmark against which the accuracy of other methods can be compared. Naturally, it has been applied extensively in quantum chemistry in particular and many-body physics more generally. However, it must be noted that such an approach is computationally expensive, and cannot be extended to arbitrarily large systems. 

Due to the immense number of parameters in the $N$-fermion wavefunction, for moderate to large $N$ in FCI calculations, where $N$ is the number of fermions in the system, it is enticing to consider whether the ground-state properties of these systems can be found using \textit{reduced} quantities. These quantities would ideally depend on fewer parameters than the $N$-fermion wavefunction. Density functional theory (DFT) is one such approach, which utilises the single-particle electron density as the fundamental variable rather than the $N$-fermion wavefunction. It has seen wide success in capturing correlation in atomic and molecular systems, which is the correlation that arises from the Coulombic interactions, as well as larger, more complicated systems such as those emerging in materials science \cite{Hohenberg1964,Kohn1965,Parr1980,Becke2014}. DFT also benefits from excellent scalability, permitting its application to large-scale systems as, generally speaking, the approach does not scale with particle number, $N$. The reduction in computational complexity attained by utilising the electron density, or, more generally, the fermion density, comes at a cost, however, as the exact energy functional for the system is replaced by an unknown energy functional of the density. This replacement brings about numerous issues concerning the accuracy and range of applicability of DFT \cite{Ruzsinszky2011}.

An alternative approach to DFT which still uses reduced quantities utilises the two-fermion reduced density matrix (2-RDM), which may be derived from the $N$-particle density matrix by integrating out the degrees of freedom of all but two of the fermions. The utility of the approach arises from the property, first demonstrated by Husimi \cite{Husimi1940}, L{\"o}wdin \cite{Lowdin1955}, and Mayer \cite{Mayer1955}, that any quantum system of $N$ particles with at-most pairwise interactions can be described completely by the 2-RDM \cite{Husimi1940}. In such a case, the variational object is the 2-RDM and the energy functional to minimise is the Hamiltonian itself. In other words, physical characteristics of the system such as the ground-state energy, particle density, and the pair-correlation function can be expressed as linear functionals of the 2-RDM. Earlier, Dirac \cite{Dirac1930} had shown that the HF ground-state for an $N$-fermion system could be expressed explicitly in terms of the one-fermion reduced density matrix (1-RDM). Soon thereafter, the idea of reducing an $N$-fermion problem to an effective two-particle problem was promoted by A. J. Coleman \cite{Coleman2000}. Early attempts by Mayer \cite{Mayer1955}, Tredgold \cite{Tredgold1957}, and Coleman \cite{Coleman1963,Coleman2000} to utilise this methodology yielded poor results, with the ground-state energy being far too low. Indeed, it was soon realised that a Rayleigh-Ritz minimisation of the ground-state energy subject to variation in the elements of the 2-RDM failed to impose sufficient constraints.

These additional constraints were termed \textit{$N$-representability conditions} \cite{Coleman1963}, and were required to ensure that a trial 2-RDM that is determined variationally corresponds to a legitimate $N$-fermion wavefunction. The search for a complete set of such conditions was known as the \textit{$N$-representability problem} and was pursed actively in electronic structure theory and many-body quantum mechanics for more than five decades. While the complete solution to the $N$-representability problem eluded the theoretical community, Coleman \cite{Coleman1963} and Garrod and Percus \cite{Garrod1964} obtained certain constraints known as the \textsf{D}, \textsf{Q}, and \textsf{G} conditions. These manifested as semidefinite constraints on matrices representing the probability distribution of two fermions, one fermion and one hole, and two holes, respectively. While the \textsf{D}, \textsf{Q}, and \textsf{Q} conditions are necessary to constrain the 2-RDM to be $N$-representable they are only sufficient in the $N=2$ fermion case and additional conditions are needed for larger systems. Further conditions were soon found by Erdahl \cite{Erdahl1978}, which are known as the \textsf{T1} and \textsf{T2} conditions, which constitute semidefinite constraints on matrices corresponding to the various distributions of three fermions and holes.

Motivated by the development of sophisticated and efficient methods in semidefinite programming (SDP) \cite{Vandenberghe1996,Gartner2012}, significant progress was made during the early 2000s to apply these $N$-representability conditions while variationally determining 2-RDMs. The development of powerful interior-point methods in the late 1990's spawned the rise of interest in the 2-RDM method as a whole and, later on, significantly more powerful boundary-point methods were developed, which greatly increased the efficiency of the implementation \cite{Mazziotti2004,Mazziotti2004a,Mazziotti2011}. Examples of systems that have been analysed using this formalism include simple atomic and molecular systems \cite{Nakata2001, Mazziotti2002,Mazziotti2004,Mazziotti2004a,Mazziotti2006,Zhao2004,Fukuda2007,Nakata2008} and quantum dots \cite{Mazziotti2008} in which correlation primarily arises from the long-range Coulombic interactions in the system and also the Hubbard model \cite{Hammond2006,Nakata2008,Verstichel2012,Anderson2013}, in which short-range interactions are present. Moreover, the RDM method has been recently applied to a large range of transition metal systems \cite{McIsaac2017,Montgomery2018} as well as strongly correlated molecules arising in both organic and inorganic chemistry \cite{Pelzer2011,Kawamura2020,Hemmatiyan2018}.

In this work, we apply the RDM method to the field of trapped, ultracold few-fermion systems exhibiting a contact interaction in both the weakly- and strongly-interacting regimes. For simplicity, we consider one-dimensional systems of net zero spin yet we permit the interactions in these systems to be repulsive and attractive. Such a system serves as an excellent medium in which to demonstrate the effectiveness of the RDM method in capturing short-range interactions. In particular, the analysis of short-range, point-like interactions with the RDM methodology permits us to assess the accuracy with which correlation in these systems is captured by the commonly utilised $N$-representability conditions which, until now, have predominately been utilised in the study of long-range, Coulombic interactions. Moreover, the strength of the contact interaction in these ultracold systems can be made arbitrarily large, thus we can consequently analyse the effectiveness of the RDM method in capturing substantial correlation in these systems in both the attractive and repulsive regimes.

Ultracold quantum gases are of immense interest as they serve as a testbed for the study of many-body quantum mechanics more generally, as they are highly controllable \cite{Bloch2008,Lewenstein2012}. Also, the intricacies of the interactions in simple ultracold gases with a small number of particles serve primarily to inform the ground-state properties and dynamics of larger, and more complicated condensed matter systems \cite{Lewenstein2007}. Thus, the application of quantum chemistry techniques to trapped ultracold few-fermion systems has allowed the treatment of larger systems where an exact diagonalisation of the Hamiltonian is completely unfeasible \cite{Grining2015,Grining2015a}. In this study, the utility and accuracy of the RDM method for highly-correlated systems is demonstrated in the ground-state calculations for simple $N=2$, $4$, $8$, $10$, and $12$ fermion systems.

The organisation of this paper is as follows. In Sec.~\ref{sec:theory} we present the general theory of the RDM approach, and demonstrate how the expectation value of a many-body Hamiltonian can be expressed directly in terms of the 2-RDM. In Sec.~\ref{sec:n_rep} we briefly outline the $N$-representability problem, and discuss the constructive solution discovered by Mazziotti \cite{Mazziotti2011} while explicitly giving the matrix representations of the well-known conditions. In Sec.~\ref{sec:imp} we describe the implementation of the RDM as an SDP. In Sec.~\ref{sec:app} we briefly discuss ultracold few-fermion systems and in Sec.~\ref{sec:result} we demonstrate the results of the RDM calculations. We conclude with a discussion on the limitations of the RDM methodology, as well as discussing the key benefits of using the RDM in the ultracold few-fermion field of research. 

\section{Reduced Density Matrix Theory}\label{sec:theory}

Consider an $N$-fermion quantum system. It is an axiomatic assumption that the Hamiltonian for this system can be expressed as the sum of one-body and two-body operators as,
\begin{equation}
\hat{\mathcal{H}}=\sum_{i=1}^N\,^{(1)}\hat{h}(i)+\frac{1}{2}\sum_{i,j=1}^N\,^{(2)}\hat{V}(i,j), \label{eq:hamiltonian_general}
\end{equation}
where $\,^{(1)}\hat{h}(i)$ is a one-body operator that acts on a single fermion, $i$, and $^{(2)}\hat{V}(i,j)$ is a two-body operator that acts on the pair of fermions, $i$ and $j$. 

Let $\ket{\Psi}$ be the (pure) state vector describing the ground-state of this system. Then, the ground-state energy of the system, $\mathcal{E}_0$, is determined from the time-independent Schr\"odinger equation \cite{Schrodinger1926}, $\hat{\mathcal{H}}\ket{\Psi}=\mathcal{E}_0\ket{\Psi}$. The $N$-fermion density operator, $\,^{(N)}\hat{\rho}$, for this system, as introduced by von Neumann \cite{vonNeumann1927}, is given by the outer product of the state vector $\ket{\Psi}$ with itself, $\,^{(N)}\hat{\rho}=\ket{\Psi}\bra{\Psi}$. Projecting this operator into coordinate space yields the $N$-fermion density matrix,
\begin{align}
&\,^{(N)}D(\bm x_1,\dots,\bm x_N;\bm x'_1,\dots,\bm x_N')\notag\\
& \hspace{60px}\coloneqq \braket{\bm x_1\dots,\bm x_N|\Psi}\braket{\Psi|\bm x'_1,\dots,\bm x_N'}\notag\\
& \hspace{60px}=\Psi(\bm x_1,\dots,\bm x_N)\,\Psi^*(\bm x'_1,\dots,\bm x_N'),
\end{align}
where $\bm x_i$ denotes the spin and spatial coordinates of the $i^{\mathrm{th}}$ fermion \footnote{The variables $\bm x_1',\dots\bm x_N'$ are to be regarded as different than the $\bm x_1,\dots\bm x_N$ variables. This is a notation which is particularly convenient when considering the expectation value of an operator wherein the density matrix appears as an integral kernel.}. The diagonal element of the $N$-fermion density matrix, when $\bm x_1'=\bm x_1,\dots,\bm x_N'=\bm x_N$, has the usual statistical interpretation: $^{(N)}D(\bm x_1,\dots,\bm x_N;\bm x_1,\dots,\bm x_N)$ is a probability distribution function describing the locations in space and spins of the $N$ fermions. 

The diagonal element of a $p$-fermion reduced density matrix ($p$-RDM) is the probability distribution function for $p<N$ fermions occupying any $p$ of the $N$ positions and spins given by $\bm x_1,\dots,\bm x_N$, and is found by integrating out the spin and spatial coordinates  of the $(p+1)$th to the $N$th fermions,
\begin{widetext}
\begin{align}
^{(p)}D(\bm x_1,\dots,\bm x_p;\bm x_1',\dots,\bm x_p')=p!\,\binom{N}{p}\int\text{d}\bm x_{p+1}\cdots\text{d}\bm x_N\,^{(N)}D(\bm x_1,\dots,\bm x_N;\bm x_1',\dots,\bm x_p',\bm x_{p+1},\dots,\bm x_N).\label{eq:prdm}
\end{align}
\end{widetext}
This object was first introduced by Husimi \cite{Husimi1940} although Dirac \cite{Dirac1930} showed in 1930 that a Hartree-Fock ground-state for a fermionic system can be expressed in terms of 1-RDMs. The normalisation factor $p!\,\binom{N}{p}$ was introduced by McWeeny \cite{McWeeny1960}. Other normalisation factors exist in the literature today, with the factor $\binom{N}{p}$ due to L\"owdin \cite{Lowdin1955} and the factor of unity due to ter Haar \cite{terHaar1961} and Coleman \cite{Coleman1963}.

A key advantage in employing $p$-RDMs lies in a fundamental result that the expectation value of a $p$-body operator can be expressed as a linear functional of the $p$-RDM \cite{Husimi1940,Lowdin1955,McWeeny1960,terHaar1961,Coleman1963}. Since the Hamiltonian in Eq.~\eqref{eq:hamiltonian_general} is the sum of one- and two-body operators, the ground-state energy, $\mathcal{E}_0=\braket{\Psi|\hat{\mathcal{H}}|\Psi}$, can be written as a linear functional of the 1- and 2-RDMs, which are given by
\begin{widetext}
\begin{align}
^{(1)}D(\bm x_1;\bm x_1')&=N\,\int\text{d}\bm x_2\cdots\text{d}\bm x_N\,^{(N)}D(\bm x_1,\dots\bm x_N;\bm x_1',\bm x_2,\dots,\bm x_N),\\
^{(2)}D(\bm x_1,\bm x_2;\bm x_1',\bm x_2')&=N(N-1)\,\int\text{d}\bm x_3\cdots\text{d}\bm x_N\,^{(N)}D(\bm x_1,\dots,\bm x_N;\bm x_1',\bm x_2',\bm x_3,\dots\bm x_N).
\end{align}
\end{widetext}
To apply the RDM apparatus to physical problems, a finite basis set must be introduced. Typically, this is a single-particle basis of spin-orbitals, which we denote by $\{\varphi_i(\bm x)\}$. Let us denote the size of the spin-orbital basis set by $K$. Due to the two-fold spin multiplicity of fermions, a rank $K$ spin-orbital basis means we have $K/2$ spatial-orbital functions in the given basis. In such a basis, the 1- and 2-RDMs are represented by
\begin{align}
^{(1)}D(\bm x_1;\bm x_1')&=\sum_{ij}\,^{(1)}D^i_j\,\varphi_i(\bm x_1)\,\varphi_j^*(\bm x_1'),\label{eq:1rdm_basis}\\
^{(2)}D(\bm x_1,\bm x_2;\bm x_1',\bm x_2')&=\sum_{ijkl}\,^{(2)}D^{ij}_{kl}\varphi_i(\bm x_1)\varphi_j(\bm x_2)\notag\\
& \hspace{40px} \times \varphi_k^*(\bm x_1')\varphi_l^*(\bm x_2').\label{eq:2rdm_basis}
\end{align}
In Eq.~\eqref{eq:1rdm_basis} and \eqref{eq:2rdm_basis} the tensors $^{(1)}D^i_j$ and $^{(2)}D^{ij}_{kl}$ can be expressed conveniently in second-quantised notation, providing the most common expressions for the 1- and 2-RDMs as found in the literature,
\begin{align}
^{(1)}D^i_j&=\braket{\Psi|\hat{a}^{\dagger}_i\hat{a}_j|\Psi},\label{eq:1rdm_second_quantised}\\
^{(2)}D^{ij}_{kl}&=\braket{\Psi|\hat{a}^{\dagger}_i\hat{a}^{\dagger}_j\hat{a}_l\hat{a}_k|\Psi},\label{eq:2rdm_second_quantised}
\end{align}
where $\hat{a}^{\dagger}$ and $\hat{a}$ are the annihilation and creation operators for fermions. We also note that we can express the general Hamiltonian in Eq.~\eqref{eq:hamiltonian_general} in a second-quantised form as
\begin{equation}
\hat{\mathcal{H}}=\sum_{ij}\,^{(1)}h^i_j\hat{a}^{\dagger}_i\hat{a}_j+\frac{1}{2}\sum_{ijkl}\,^{(2)}V^{ij}_{kl}\,\hat{a}^{\dagger}_i\hat{a}^{\dagger}_j\hat{a}_l\hat{a}_k,\label{eq:hamiltonian_second_quantised}
\end{equation}
where the indices $i,j,k$ and $l$ correspond to elements of the spin-orbital basis set $\{\varphi_i(\bm x)\}$, the sums run over all possible values of the indices in the basis, $i,j,k,l\in\{1,2,\dots,K\}$, and $^{(1)}\hat{h}^i_j$ and $^{(2)}\hat{V}^{ij}_{kl}$ are the one- and two-fermion integrals,
\begin{align}
^{(1)}h^i_j&=\braket{i|\,^{(1)}\hat{h}|j}=\int\text{d}\bm x\,\varphi_i^*(\bm x)\,^{(1)}\hat{h}\,\varphi_j(\bm x),\\
^{(2)}V^{ij}_{kl}&=\braket{ij|\,^{(2)}\hat{V}|kl}=\int\text{d}\bm x_1\,\text{d}\bm x_2\,\varphi_i^*(\bm x_1)\varphi_j^*(\bm x_2)\notag\\
&\hspace{75px}  \times \,^{(2)}\hat{V}\,\varphi_k(\bm x_1)\varphi_l(\bm x_2). 
\end{align}
By taking the expectation value of the Hamiltonian in Eq.~\eqref{eq:hamiltonian_second_quantised} with respect to the ground-state $\ket{\Psi}$ and incorporating the expressions of the 1- and 2-RDMs in second-quantised notation (Eqs.~\eqref{eq:1rdm_second_quantised} and \eqref{eq:2rdm_second_quantised}) we see that the ground-state energy is a linear functional of the 1- and 2-RDMs;
\begin{align}
\braket{\Psi|\hat{\mathcal{H}}|\Psi}&=\sum_{ij}\,^{(1)}h^i_j\,^{(1)}D^i_j+\frac{1}{2}\sum_{ijkl}\,^{(2)}V^{ij}_{kl}\,^{(2)}D^{ij}_{kl}\notag\\
&=\mathsf{Tr}\left\{\,^{(1)}h^i_j\,^{(1)}D^i_j\right\}+\frac{1}{2}\mathsf{Tr}\left\{\,^{(2)}V^{ij}_{kl}\,^{(2)}D^{ij}_{kl}\right\}.\label{eq:energy_rdm}
\end{align}

\section{$N$-Representability}\label{sec:n_rep}

Initial efforts to calculate the ground-state properties of simple systems with the RDM method yielded drastically incorrect results \cite{Coleman1963,Tredgold1957,Ayres1958,Garrod1964,Coleman2000}. This is due to the lack of $N$-representability conditions applied to the 2-RDM. The search for these $N$-representability conditions, which ensure that the 1- and 2-RDMs correspond to a legitimate $N$-particle wavefunction, are what Coleman \cite{Coleman1963} called the \textit{$N$-representability problem}.

We denote by $^{(N)}\mathcal{P}$ the set of all positive semidefinite Hermitian operators of unit trace on the Hilbert space for our $N$-fermion system. This set is equivalent to the set of all possible $N$-fermion density operators $^{(N)}D$ \cite{Coleman1963,vonNeumann2018}. We also denote by $^{(p)}\mathcal{P}_N$ the set of all $N$-representable $p$-RDMs. The prescription given in Eq.~\eqref{eq:prdm} permits one to calculate a $p$-RDM from the $N$-particle density matrix but does not provide a mechanism to ensure that an alleged $p$-RDM corresponds to a legitimate $N$-particle density matrix. That is, a variationally determined $p$-RDM does not necessarily belong to the set $^{(p)}\mathcal{P}_N$ but to the more general set of $p$-RDMs which may not be $N$-representable, the set of which we denote $^{(p)}\mathcal{P}$. Hence, $^{(p)}\mathcal{P}_N\subset\,^{(p)}\mathcal{P}$, i.e., $^{(p)}\mathcal{P}_N$ is a \textit{proper} subset of $^{(p)}\mathcal{P}$. Thus, the $N$-representability problem can be stated as the search for a complete characterisation of the set $^{(p)}\mathcal{P}_N$ as a subset of $^{(p)}\mathcal{P}$. 

Now, the set $^{(N)}\mathcal{P}$ is convex as are its subsets $^{(p)}\mathcal{P}$ and $^{(p)}\mathcal{P}_N$ , with $^{(p)}\mathcal{P}_N$ being a convex subset of $^{(p)}\mathcal{P}$. A convex set is a set $\mathcal{S}$ such that if $x,y\in\mathcal{S}$ then $\alpha x+\beta y\in \mathcal{S}$, for all $\alpha,\beta\in \mathbbm{R}_{\ge 0}$ such that $\alpha+\beta=1$. That is, convex combinations of density matrices are themselves density matrices. An extreme element of a convex set is an element that cannot be expressed as a convex combination of other elements in the set \cite{Rockafellar1972,HiriartUrrut2004}. These are the density matrices representing pure states as opposed to mixed states, which are themselves represented by convex combinations of pure states. By the Krein-Milman theorem \cite{Krein1940}, a convex set is completely specified by its extreme elements. From this theorem, a complete characterisation of the set $^{(p)}\mathcal{P}_N$ can be garnered from its extreme elements, i.e., by considering the pure state density matrices of the system.

The complete characterisation was first discovered by Mazziotti \cite{Mazziotti2012,Mazziotti2012a}. Define the convex set of $p$-fermion operators that are positive semidefinite in their trace against an $N$-representable $p$-RDM by $^{(p)}\mathcal{P}_N^*$,
\begin{align*}
^{(p)}\mathcal{P}_N^*\coloneqq \left\{\,^{(p)}\hat{O}\,|\mathsf{tr}(\,^{(p)}D\,^{(p)}\hat{O})\ge 0, \ \forall\,^{(p)}D\in\,^{(p)}\mathcal{P}_N\right\}.
\end{align*}
Kummer \cite{Kummer1967} showed that this set explicitly exists, and this set provides a complete characterisation of its dual or polar set, i.e., the set of $N$-representable $p$-RDMs, $^{(p)}\mathcal{P}_N$. As such, a knowledge of the extreme elements in $^{(p)}\mathcal{P}_N^*$ allows us to completely characterise the set of $p$-RDMs. Mazziotti noted the important relation that $^{(2)}\mathcal{P}_N^*\subseteq \,^{(3)}\mathcal{P}_N^*\subseteq\cdots\subseteq \,^{(K)}\mathcal{P}_N^*,$ where $K$ is the rank of the spin orbital basis. As such, the extreme elements of $^{(2)}\mathcal{P}_N^*$ are specified by convex combinations of the extreme elements of $^{(K)}\mathcal{P}_N^*$, which are of the form $^{(K)}\hat{O}_i=\,^{(K)}\hat{C}_i\,^{(K)}\hat{C}_i^{\dagger}$, where $^{(K)}\hat{C}_i$ is a polynomial in creation and annihilation operators of order $\le K$. Hence,  the operators which constrain the $2$-RDM to be $N$-representable, for example, are represented by $^{(2)}\hat{O}=\sum_i\omega_i\,\hat{C}_i\,\hat{C}_i^{\dagger}$ where the weights $\omega_i$ are chosen such that all three-body or higher operators cancel on their trace against the 2-RDM. 

Such a solution yields a collection of $(p,q)$-positivity conditions \cite{Mazziotti2011}, where the $p$ indicates the highest RDM required to evaluate the condition (here that remains the 1-, and 2-RDM) and the $q$ indicates the highest $q$-body operator cancelled by the combinations expressed above. For example, the $(1,1)$-positivity conditions are the conditions on the 1-RDM which come from considering operators $\hat{C}_D=\sum_jb_j\,\hat{a}_j^{\dagger}$ and  $\hat{C}_Q=\sum_jb_j\,\hat{a}_j$. Keeping the trace of these operators with the ground-state wavefunction greater than or equal to zero yields the $N$-representability conditions on the 1-RDM, first found by Coleman \cite{Coleman1963};
\begin{align}
\braket{\Psi|\hat{a}^{\dagger}_i\hat{a}_j|\Psi}=\,^{(1)}D^i_j&\succcurlyeq 0\label{eq:nrep_1rdma}\\
\braket{\Psi|\hat{a}_i\hat{a}_j^{\dagger}|\Psi}=\delta^i_j-\,^{(1)}D^i_j&\succcurlyeq 0,\label{eq:nrep_1rdmb}
\end{align}
where $\succcurlyeq 0$ means `positive semiefinite', $\delta$ is the Kronecker delta, and Eq.~\eqref{eq:nrep_1rdmb} follows from the anticommutation relation for fermion creation and annihilation operators. The constraints here are necessary and complete, that is, the 1-RDM is completely $N$-representable if equipped with these constraints, and the constraints are equivalent to the Pauli exclusion principle: the occupation numbers of a given state in a fermion system must be 0 or 1 \cite{Coleman1963}.

The $(2,2)$-positivity conditions arise from considering the operators $\hat{C}_D=\sum_{ij}b_{ij}\,\hat{a}^{\dagger}_j\,\hat{a}_k^{\dagger}$, $\hat{C}_Q=\sum_{ij}b_{ij}\,\hat{a}_j\hat{a}_k$, and $\hat{C}_G=\sum_{ij}b_{ij}\hat{a}_k^{\dagger}\hat{a}_k$. Restricting the trace of the operators $^{(2)}\hat{O}_D, ^{(2)}\hat{O}_Q$, and $^{(2)}\hat{O}_G$ then gives us the well-known \textsf{D, Q}, and \textsf{G} conditions, first due to Coleman \cite{Coleman1963} and Garrod and Percus \cite{Garrod1964},
\begin{align}
\braket{\Psi|\hat{a}_i^{\dagger}\hat{a}_j^{\dagger}\hat{a}_l\hat{a}_k|\Psi}=\,^{(2)}D^{ij}_{kl}&\succcurlyeq0\label{eq:nrep_2rdma}\\
\braket{\Psi|\hat{a}_i\hat{a}_j\hat{a}_l^{\dagger}\hat{a}_k^{\dagger}|\Psi}=\,^{(2)}Q^{ij}_{kl}&\succcurlyeq0\label{eq:nrep_2rdmb}\\
\braket{\Psi|\hat{a}_i^{\dagger}\hat{a}_j\hat{a}_l^{\dagger}\hat{a}_k|\Psi}=\,^{(2)}G^{ij}_{kl}&\succcurlyeq0\label{eq:nrep_2rdmc}.
\end{align}
The conditions given in Eq.~\eqref{eq:nrep_2rdma}, Eq.~\eqref{eq:nrep_2rdmb}, and Eq.~\eqref{eq:nrep_2rdmc} are physically interpreted as constraining the probability distributions of two fermions, two holes, and one fermion and one hole to be non-negative, respectively. They do not form a complete set, however, of $N$-representability conditions on the 2-RDM. For example, the $(2,3)$-positivity conditions are also necessary for a description of a system of $N>2$ fermions. Combinations of constraint matrices derived from considering $\hat{C}_i$ operators of degree 3 are utilised, since these can be expressed as linear combinations of the elements of the 1- and 2-RDMs. These are called the \textsf{T1} and \textsf{T2} conditions, and have the expression \cite{Erdahl1978},
\begin{align}
^{(3)}T1=\,^{(3)}D+\,^{(3)}Q&\succcurlyeq0\label{eq:nrep_t1}\\
^{(3)}T2=\,^{(3)}E+\,^{(3)}F&\succcurlyeq0\label{eq:nrep_t2},
\end{align}
where
\begin{align}
^{(3)}D^{ijk}_{lmn}&=\braket{\Psi|\hat{a}_i^{\dagger}\hat{a}_j^{\dagger}\hat{a}_k^{\dagger}\hat{a}_n\hat{a}_m\hat{a}_l|\Psi}\label{eq:nrep_2rdmd}\\
^{(3)}E^{ijk}_{lmn}&=\braket{\Psi|\hat{a}_i^{\dagger}\hat{a}_j^{\dagger}\hat{a}_k\hat{a}_n^{\dagger}\hat{a}_m\hat{a}_l|\Psi}\label{eq:nrep_2rdme}\\
^{(3)}F^{ijk}_{lmn}&=\braket{\Psi|\hat{a}_l\hat{a}_m\hat{a}_n^{\dagger}\hat{a}_k\hat{a}^{\dagger}_j\hat{a}^{\dagger}_i|\Psi}\label{eq:nrep_2rdmf}\\
^{(3)}Q^{ijk}_{lmn}&=\braket{\Psi|\hat{a}_l\hat{a}_m\hat{a}_n\hat{a}_k^{\dagger}\hat{a}_j^{\dagger}\hat{a}_i^{\dagger}|\Psi}\label{eq:nrep_2rdmg}.
\end{align}
These four matrices corresponding to the distributions of three fermions, two fermions and a hole, one fermion and two holes, and three holes, respectively. Although these matrices are created through consideration of three-body operators, the combinations given by Eq.~\eqref{eq:nrep_t1} and Eq.~\eqref{eq:nrep_t2} constrain the 2-RDM through cancellation. The \textsf{T1} and \textsf{T2} conditions were found implicitly by Erdahl \cite{Erdahl1978} and first implemented by Zhao \cite{Zhao2004} and Mazziotti \cite{Mazziotti2002,Mazziotti2006}.

All conditions expressed here can be recast as linear constraints on the 1- and 2-RDMs by implementing the fermion anti-commutation relations. For example, the $^{(2)}Q$ and $^{(2)}G$ matrices can be expressed as,
\begin{align}
^{(2)}Q^{ij}_{kl}&=\delta^i_k\delta^j_l-\delta^i_l\delta^j_k-\delta^i_j\,^{(1)}D^j_l-\delta^j_l\,^{(1)}D^i_j+\delta^i_l\,^{(1)}D^j_k\notag\\
& \hspace{10px} +\delta^j_k\,^{(1)}D^i_l+2\,^{(2)}D^{ij}_{kl},\\
^{(2)}G^{ij}_{kl}&=\delta^i_k\,^{(1)}D^i_j-2\,^{(2)}D^{il}_{kj}.
\end{align}
Similar expressions can be found for the $^{(3)}T1$ and $^{(3)}T2$ matrices are given elsewhere (for example, see Zhao \textit{et al.} \cite{Zhao2004}). With these expressions, a set of linear constraints on the components of the 1- and 2-RDMs are now available in a variational calculation. In this work, we include conditions up to and including the \textsf{T1} and \textsf{T2} conditions and therefore, except in the $N=2$ fermion case, we hold the 2-RDM to be be \textit{approximately} $N$-representable. We will see that the $(2,2)$- and $(2,3)$-positivity conditions, given by constraining $^{(2)}D, ^{(2)}Q,$ and $^{(2)}G$, and $^{(3)}T1$ and $^{(3)}T2$ to be positive semidefinite, yield accurate results for the ground-state energy at a range of short-range interaction strengths when compared to other methods.

\subsection{Spin-Adaption}

The basis set we introduced is such that each spin-orbital is a product of a spatial wavefunction and a number representing the spin, $\varphi_i(\bm x)=\phi_i(\bm r)\sigma_i(\omega)$ where $\sigma_i(\omega)=\alpha(\omega)$ or $\sigma_i(\omega)=\beta(\omega)$ if spin-up or spin-down, respectively, where $\alpha$ and $\beta$ are eigenfunctions of the spin operator $\hat{\mathcal{S}}_z$. That is, we have $L$ spatial wavefunctions and hence $K=2L$ spin-orbitals, allowing for the twofold spin multiplicity. We also order our basis appropriately such that the first $L$ spin-orbitals all are spin-up (or all spin-down) and the next $L$ are all spin-down (or all spin-up). Then, the spin-orbitals indicated by the index 1 and $L+1$ have the same spatial wavefunction, but differ in their spin.

Through this construction of the basis, we render our 2-RDM and associated matrices block-diagonal in a process known as \textit{spin-adaption}. Typically, the term spin-adaption refers to constraining the basis functions to be simultaneous eigenfunctions of $\hat{\mathcal{S}}_z$ \textit{and} $\hat{\mathcal{S}}^2$ \cite{Szabo1996}, but here we simply enforce the former condition and implement the latter as a constraint enforced during the variational optimisation through Eq.~\eqref{eq:spin_constraint}. This form of implementation has been used numerous times in applications to atomic and molecular systems \cite{Nakata2001,Mazziotti2002,Juhasz2004,Zhao2004,Fukuda2007,Nakata2008}. Spin and symmetry adaption from both the $\hat{\mathcal{S}}_z$ and $\hat{\mathcal{S}}^2$ operator of the $^{(2)}D, ^{(2)}Q,$ and $^{(2)}G$ matrices has been analysed \cite{Gidofalvi2005}. 

The spin-adapted 1-RDM has the form,
\begin{align}
^{(1)}D^i_j=\begin{pmatrix}
^{(1)}D^{\alpha}_{\alpha} & \\
& ^{(1)}D^{\beta}_{\beta}
\end{pmatrix}
\end{align}
since in our basis $^{(1)}D^{\alpha}_{\beta}=\,^{(1)}D^{\beta}_{\alpha}=0$. Here, the blocks are identical due to the ordering of the spin-orbital basis, and have size $(K/2)\times (K/2)$. The spin-adapted 2-RDM has the form,
\begin{align}
^{(2)}D^{ij}_{kl}=\begin{pmatrix}
^{(2)}D^{\alpha\alpha}_{\alpha\alpha} & & \\
& ^{(2)}D^{\beta\beta}_{\beta\beta} & \\
& & ^{(2)}D^{\alpha\beta}_{\alpha\beta}
\end{pmatrix}.
\end{align}
In the case of a closed-shell system, where each spatial orbital is doubly-occupied, the first and second blocks of the 1- and 2-RDMs are identical, and the sizes of the first two blocks of the 2-RDM are $K(K/2-1)/4\times K(K/2-1)/4 $, whereas the third block has size $K^2/4$. The $^{(2)}Q$ matrix has the same block structure as the 2-RDM since it is the two-hole reduced density matrix and holes obey the same statistics. The one-hole one-fermion RDM, $^{(2)}G$, has a more complicated block structure as it does not exhibit as many symmetry properties as the other matrices. It has the spin-adapted form,
\begin{equation}
^{(2)}G^{ij}_{kl}=\begin{pmatrix}
^{(2)}G^{\alpha\alpha}_{\alpha\alpha} & ^{(2)}G^{\alpha\alpha}_{\beta\beta} & &\\
^{(2)}G^{\beta\beta}_{\alpha\alpha} & ^{(2)}G^{\beta\beta}_{\beta\beta} & &\\
& & ^{(2)}G^{\alpha\beta}_{\alpha\beta} & \\
& & & ^{(2)}G^{\beta\alpha}_{\beta\alpha}
\end{pmatrix}.
\end{equation}
The block structure of the $^{(3)}T1$ and $^{(3)}T2$ matrices have more complicated representation which we will not give here (see Zhao \textit{et al}, \cite{Zhao2004}). 

\subsection{A Summary of the $N$-Representability Conditions}

In addition to the $N$-representability conditions which manifest in the \textsf{D, Q, G, T1}, and \textsf{T2} conditions, we have the conditions that hold for density matrices in general. We list these conditions here.
\begin{itemize}
\item[\textsf{i)}] The 2-RDM is anti-symmetric in its upper and lower indices,
\begin{align}
^{(2)}D^{ij}_{kl}&=-\,^{(2)}D^{ji}_{kl}\notag\\
&=-\,^{(2)}D^{ij}_{lk}\notag\\
&=\,^{(2)}D^{ji}_{lk}.
\end{align}
The $^{(2)}Q$ matrix has exactly the same anti-symmetry as the 2-RDM, whereas the $^{(2)}G$ matrix does not. Furthermore, the $^{(3)}T1$ matrix is anti-symmetric with respect to interchange of its triples of indices, and the $^{(3)}T2$ matrix is anti-symmetric with interchange of the latter pairs in each triple. 
\item[\textsf{ii)}] Hermiticity of the 1- and 2-RDMs,
\begin{align}
^{(1)}D^i_j&=\left(\,^{(1)}D^j_i\right)^*,\\
^{(2)}D^{ij}_{kl}&=\left(^{(2)}D^{kl}_{ij}\right)^*.
\end{align}
This Hermiticity is also extended to the $^{(2)}Q,\,^{(2)}G,\,^{(3)}T1$ and $^{(3)}T2$ matrices. 
\item[\textsf{iii)}] We ensure that the constant number of fermions (spin-up and spin-down) in the physical system is reflected in the trace of the 1- and 2-RDMs. That is,
\begin{align}
\mathsf{tr}\left\{\,^{(1)}D^i_j\right\}&=\sum_i\,^{(1)}D^i_i=N,\\
\mathsf{tr}\left\{\,^{(2)}D^{ij}_{kl}\right\}&=\sum_{ij}\,^{(2)}D^{ij}_{ij}=N(N-1).
\end{align}
We recall that this normalisation scheme is a convention, as the 1- and 2-RDMs can be normalised to unity.
\item[\textsf{iv)}] The number of spin-up (or spin-down) fermions is held constant, a constraint which must show up in the normalisation:
\begin{align}
\mathsf{tr}\left\{\,^{(1)}D^{i_{\alpha}}_{j_{\alpha}}\right\}&=\sum_{i_{\alpha}}\,^{(1)}D^{i_{\alpha}}_{i_{\alpha}}=N_{\alpha},\\
\mathsf{tr}\left\{\,^{(2)}D^{i_{\alpha}j_{\alpha}}_{{k_{\alpha}l_{\alpha}}}\right\}&=\sum_{i_{\alpha}j_{\alpha}}\,^{(2)}D^{i_{\alpha}j_{\alpha}}_{i_{\alpha}j_{\alpha}}\notag\\
&=N_{\alpha}(N_{\alpha}-1),
\end{align}
where $i_{\alpha},j_{\alpha},\dots$ indicate indices corresponding to spin-up orbitals, and $N_{\alpha}$ is the number of spin-up orbitals. The latter of these constraints can be derived from the fact that the wavefunction, $\ket{\Psi}$, of the system is an eigenstate of the number operator for the spin-up electrons, $\hat{\mathcal{N}}_{\alpha}$ \cite{Nakata2001}. 

\item[\textsf{v)}] We ensure we have a linear constraint involving the net spin, $S$. This is due to the fact that $\ket{\Psi}$ is held to be an eigenfunction of the $\hat{\mathcal{S}}^2$ operator. This condition is difficult to express in the notation above, so we use an alternate notation to ensure this constraint is explicit. Following Zhao \textit{et al.} \cite{Zhao2004}, we let $n_i,n_j,\dots$ denote spatial orbital indices, with $n_i\alpha$ and $n_i\beta$ denoting spin-up and spin-down spatial orbitals, respectively.

Then, we ensure
\begin{align}
&\sum_{n_i,n_j}\left\{\,^{(2)}D^{n_i\alpha\,n_j\alpha}_{n_i\alpha\,n_j\alpha}+\,^{(2)}D^{n_i\beta\,n_j\beta}_{n_i\beta\,n_j\beta}\right\}\notag\\
& \ \ \ -2\sum_{n_i,n_j}\,^{(2)}D^{n_i\alpha\,n_j\beta}_{n_i\alpha\,n_j\beta}\notag\\
& \ \ \ -4\sum_{n_i,n_j}\,^{(2)}D^{n_i\alpha\,n_j\beta}_{n_j\alpha\,n_i\beta}+3N=4S(S+1).  \label{eq:spin_constraint}
\end{align}
Note that in all applications in this work, the net spin is always zero, i.e., $S=0$.
\item[\textsf{vi)}] We ensure we have the correct spin-symmetries of the matrices $^{(1)}D,\,\mathbbm{1}-\,^{(1)}D,\,^{(2)}D,\,^{(2)}Q,\,^{(2)}G,\,^{(3)}T1$, and $^{(3)}T2$. For example, we ensure
\begin{align}
^{(2)}D^{i_{\sigma_i}j_{\sigma_j}}_{k_{\sigma_k}l_{\sigma_l}}=0, \ \ \text{when }\sigma_i+\sigma_j\neq \sigma_k+\sigma_l,
\end{align}
where $\sigma_i$ is the spin of the $i$th spin-orbital. There are similar constraints for the remaining matrices, which are listed elsewhere \cite{Zhao2004}.  
\item[\textsf{vii)}] The $N$-representability constraints; the positive semidefiniteness of the matrices $^{(1)}D,\,\mathbbm{1}-\,^{(1)}D,\,^{(2)}D,\,^{(2)}Q,\,^{(2)}G,\,^{(3)}T1$, and $^{(3)}T2$. 
\end{itemize}
Due to the symmetries (and anti-symmetries) of the Hamiltonian and 1- and 2-RDMs, as well as the other matrices that are constrained to be positive semidefinite, we can restrict the elements of the matrices we use in the calculation to increase efficiency. For example, due to the anti-symmetry of the 2-RDM, we can now only consider elements of $^{(2)}D^{ij}_{kl}$ for $1\le i<j\le K$ and $1\le k<l\le K$. We can do the same for $^{(2)}Q^{ij}_{kl}$. However, the $^{(2)}G$ matrix does not permit such a reduction as it does not have the same anti-symmetric structure. For $^{(3)}T1^{ijk}_{lmn}$ we can take $1\le i<j<k\le K$ and $1\le l<m<n\le K$, and for $^{(3)}T2^{ijk}_{lmn}$ we take those elements with $1\le j<k\le K$ and $1\le n<m\le K$, where $K$ is the spin-orbital basis rank. 

Similarly, in the Hamiltonian given in Eq.~\eqref{eq:hamiltonian_second_quantised}, with the matrix elements of the two-body operator, $^{(2)}V^{ij}_{kl}$, we take the elements $1\le i<j\le K$ and $1\le k<l\le K$. As such, the optimisation problem becomes, 
\begin{align}
\braket{\Psi|\hat{\mathcal{H}}|\Psi}=\sum_{ij}\,^{(1)}h^i_j\,^{(1)}D^i_j+\sum_{i<j}\sum_{k<l}\,^{(2)}\tilde{V}^{ij}_{kl}\,^{(2)}D^{ij}_{kl},\label{eq:energy_unique}
\end{align}
where $^{(2)}\tilde{V}$ is now the anti-symmetrised two-fermion integral,
\begin{align}
^{(2)}\tilde{V}^{ij}_{kl}=\braket{ij|\,^{(2)}\hat{V}|kl}-\braket{ij|\,^{(2)}\hat{V}|lk}.
\end{align}

\section{Implementation as a Semidefinite Program}\label{sec:imp}

The type of optimisation problem given by Eq.~\eqref{eq:energy_unique} with the necessity to constrain the constituent matrices to be positive semidefinite is one that naturally facilitates the use of semidefinite programming. A semidefinite program (SDP) is a general optimisation problem that can be expressed as \cite{Vandenberghe1996,Gartner2012},
\begin{equation}
\begin{cases}
\min\limits_y & h^Ty\\
\text{subject to} & \displaystyle\sum_{p=1}^m\,\mathsf{A}_p\,y_p-\mathsf{C}\succcurlyeq0,
\end{cases}
\end{equation}
where $h,y\in\mathbbm{R}^m$ are vectors (where $m$ is the number of constraints), and $\mathsf{A}_p, \mathsf{C}\in\mathbbm{S}^n$ are real symmetric matrices of size $n$ (where $n$ depends on the number of constraints and the basis rank. See Fukuda \textit{et al}, \cite{Fukuda2007}). Here, $h^T$ denotes the transpose of $h$, such that the product $h^Ty$ represents the typical inner product between two Euclidean vectors. 

We formulate the RDM variational problem as an SDP by letting the vector $h$ contain the matrix elements of the Hamiltonian and the vector $y$ contain the components of the 1- and 2-RDM which will be determined variationally. Then, by judicious choices of the matrices $\mathsf{A}_p$ and $\mathsf{C}$ we can implement the conditions (1)-(7) given in the previous section. Explicitly, let us define a linear transformation: $\mathsf{svec}:\mathbbm{S}^n\to \mathbbm{R}^{n(n+1)/2}$, where $\mathbbm{S}^n$ is the set of real, symmetric matrices of size $n\times n$, by
\begin{align*}
\mathsf{svec(U)}=(U_{11},\sqrt{2}U_{12},\sqrt{2}U_{13},\dots,U_{22},\sqrt{2}U_{23},\dots)^T,
\end{align*}
where $\mathsf{U}\in\mathbbm{S}^n$. The appearance of the $\sqrt{2}$ ensures that we need not consider two products of the form, for example, $^{(1)}h^1_2\,^{(1)}D^1_2$ and $^{(1)}h^2_1\,^{(1)}D^2_1$, but simply double-count \textit{one} of them as they are equal. This transformation allows us to consistently place the elements of a symmetric matrix into a vectorised form. Hence, in the SDP above, we let $h=(\mathsf{svec}(\,^{(1)}h^i_j)^T,\mathsf{svec}(\,^{(2)}\tilde{V}^{ij}_{kl})^T)^T$ and similarly for the $y$ vector which contains elements of the 1- and 2-RDMs \cite{Zhao2004,Fukuda2007}. We then ensure that the matrix held to be positive semidefinite, which is constructed by the $\mathsf{A}_p$ and $\mathsf{C}$ matrices, has the following diagonal blocks: $^{(1)}D,\,^{(2)}D,\,^{(2)}Q,\,^{(2)}G,\,^{(3)}T1,$ and $^{(3)}T2$, i.e.,
\begin{align*}
\sum_{p=1}^m\mathsf{A}_py_p-\mathsf{C}=\begin{pmatrix}
^{(1)}D & & & &\\
& \mathbbm{1}-\,^{(1)}D & & &  \\
& & ^{(2)}D  & & \\
& & & ^{(2)}Q  & \\
& & & & \ddots
\end{pmatrix}
\end{align*}
Hence, if we consider the first block in which we are constraining  $^{(1)}D$ to be positive semidefinite, we note that the first elements of the $y$ matrix are those containing the elements of $^{(1)}D$, i.e., $y=(\,^{(1)}D^1_1,\sqrt{2}\,^{(1)}D^1_2,\dots)$ then one can see that in the corresponding blocks of $\mathsf{A}_p$, we would have
\begin{align*}
\mathsf{A}_1^{\text{1st block}}=\begin{pmatrix}
1 & 0 & \cdots\\
0 & 0 & \cdots\\
\vdots & \vdots & \ddots
\end{pmatrix}, \ \ \mathsf{A}_2^{\text{1st block}}=\begin{pmatrix}
0 & 1 & 0 & \cdots\\
0 & 0 & 0 & \cdots\\
\vdots & \vdots & \vdots & \ddots
\end{pmatrix}
\end{align*}
and the first block of the $\mathsf{C}$ matrix would contain all zeros. As such, the matrices involved in such a calculation are extremely sparse. The size of sparsity of the SDPs for RDM implementation has been covered in detail elsewhere \cite{Fukuda2007}.

The formulation of the RDM apparatus as an SDP was implicit in the early works of Garrod \textit{et al.} \cite{Garrod1975,Garrod1976} on atomic beryllium and Rosina \textit{et al.} \cite{Rosina1975}. The development of powerful \textit{interior points methods} \cite{Helmberg1994,Wright1997} for SDPs were developed largely in the late 1990s which led to a resurgence in the application of the RDM theory to atoms and molecules \cite{Nakata2001,Mazziotti2002,Zhao2004}. These calculations yielded accurate results, but were restricted to relatively small basis sizes. However, \textit{boundary-point methods} \cite{Povh2006} have facilitated an enormous increase in computational efficiency, allowing the analysis of more complicated systems \cite{Mazziotti2004,Mazziotti2004a,Mazziotti2011}. In this work, we utilise the Semidefinite Programming Algorithm (SDPA) \cite{Yamashita2010,Yamashita2012} which utilises interior point methods, as well as SDPNAL$+$ \cite{Zhao2010,Yang2015,Defeng2020} which utilises an augmented Lagrangian method \cite{Zhao2010,Sun2008}, rendering it extremely efficient for large-scale SDPs. To implement these programs, we utilised Spartan high-performance computing system at the University of Melbourne \cite{Meade2017}.

\section{Application to Ultracold Few-Fermion Systems}\label{sec:app}

In this work, the RDM method is not spruiked as a method leading to extraordinary accuracy relative to available methods in the field, but a promising alternative to treat large systems that is competitive in its accuracy and efficiency to known methods \textit{whatever} the basis used. Due to the method's dependence on the basis rank and \textit{not} the particle number $N$, improvements in optimisation will allow small to moderately sized systems that are highly correlated to be analysed. 

The utility of the RDM method has been demonstrated in the field of quantum chemistry but has seen few applications outside this field. Here, we apply the method in another field, that of ultracold few-fermion systems. We demonstrate that the RDM method with no more constraints than the \textsf{T1} and \textsf{T2}, which are widely implemented in quantum chemistry, are sufficient to capture strongly correlated ultracold systems exhibiting contact interactions even as the system approaches the unitary regime, where the interaction strength becomes arbitrarily large.

We choose the most simple of such systems: a one-dimensional, harmonically trapped system of $N$ particles with even populations of spin-up and spin-down particles with a short-range $s$-wave interaction. These systems serve as a convenient medium to assess the accuracy of the RDM method, and the utilisation of the \textsf{T1} and \textsf{T2} conditions, in capturing strong correlation in general ultracold systems, as the interaction strength can be made arbitrarily large, and the intrinsic nature of the point-like contact interaction exhibited in this system is not peculiar to one-dimensional systems, permitting us to expect comparable accuracy in other dimensions. Such systems have been studied at length both theoretically \cite{Tonks1936,Girardeau1960,Busch1998,Blume2007,vonStecher2008,Blume2009,Rubeni2012,Sowinksi2013,Gharashi2013,Grining2015,Grining2015a,Pecak2017,Sowinski2019,Bloch2008,Giorgini2008,Guan2013} and experimentally \cite{Kinoshita2004,Haller2009,Guan2013,Zurn2012,Wenz2013,Murmann2015,Zurn2013}. Naturally, quasi-one-dimensional systems are constructed by the means of relatively strong harmonic confinement in two orthogonal directions and a relatively weak confinement in a third. If the confinement in the planar direction(s) is strong enough, we can model the system with a one-dimensional Hamiltonian.

By a comparison to Eq.~\eqref{eq:hamiltonian_general}, we make the identifications,
\begin{align}
\sum_{i=1}^N\,^{(1)}\hat{h}=-\frac{1}{2}\sum_{i=1}^N\,\frac{\partial^2}{\partial x_i^2}+\frac{1}{2}\sum_{i=1}^Nx_i^2,\\
\sum_{i,j=1}^N\,^{(2)}\hat{V}=\,g_{\mathsf{1D}}\sum_{i,j=1}^M\delta(x_i-x_j),
\end{align}
Here, we are working in units where $\hbar=m=\omega=1$ where $m$ is the mass of the fermion and $\omega$ is the one-dimensional trapping frequency. Also, $g_{\mathsf{1D}}$ parametrises the interaction strength and has the units $\hbar\omega a_{\mathsf{ho}}$ where $a_{\mathsf{ho}}$ is the oscillator length of the harmonic trapping potential. The parameter $g_{\mathsf{1D}}$ is related to an effective one-dimensional scattering length $a_{\mathsf{1D}}$ via $g_{\mathsf{1D}}=-2\hbar^2/(ma_{\mathsf{1D}})$. This one-dimensional scattering length can be derived from the $s$-wave scattering length, $a_s$, and the characteristic length of the planar trapping potential $a_{\perp}$ through the equation
\begin{align}
a_{\mathsf{1D}}=-\frac{a_{\perp}^2}{2a_s}\left(1-C\frac{a_s}{a_{\perp}}\right),
\end{align}
where $C=1.4603$ \cite{Olshanii1998}. Thus, while the parameter $g_{\mathsf{1D}}$ serves as a measure of the strength of the contact interactions in our system, it is also dependent on the $s$-wave scattering length, $a_s$, and the trap geometry. The energy of this system is consequently measured in units of $\hbar\omega$. 

A convenient basis to study this system is the set of one-dimensional quantum harmonic oscillator eigenfunctions which are defined by,
\begin{equation}
\varphi_n(x)\coloneqq \frac{1}{\pi^{1/4}}\frac{1}{\sqrt{2^nn!}}\exp\left\{-\frac{x^2}{2}\right\}\,H_n(x),\label{eq:basis}
\end{equation}
where $H_n(x)$ is a Hermite polynomial. The error in many-body calculations using this basis scales approximately as $1/\sqrt{(K/2)}$ \cite{Grining2015} where we recall $K$ is the spin-orbital basis rank. Other basis sets exhibit superior scaling, such as the plane-wave basis which scales as $1/(K/2)$ \cite{Jeszenszki2020}. However, we utilise the single-particle wavefunctions in Eq.~\eqref{eq:basis} due to the simplicity of the implementation (in the harmonic oscillator basis, the matrix elements of the Hamiltonian can be calculated quickly, as the one-electron integrals are given by the well-known harmonic oscillator energy $\braket{i|\,^{(1)}\hat{h}|j}=\delta_{ij}(i+1/2)$, and the two-electron integrals can be calculated exactly using Gauss-Hermite quadrature) and to allow this paper to be easily compared to works with similar result utilising the same basis (see, for example, \cite{Sowinksi2013,Grining2015,Grining2015a,Pecak2017}).

With this choice of basis, the results presented here do not exhibit any improvement in the accuracy of the ground-state energies of $N$ fermion systems relative to results already presented in the literature, yet the results do show that the RDM method with appropriate $N$-representability conditions accurately captures the correlation in these systems when compared to other well-known, and accurate, methods. 

To assess the accuracy of the ground-state energy found from the RDM method, we compare it to the exact ground-state energy of the system, in the given basis, if feasible. This is found from the FCI method, which is equivalent to a direct diagonalisation of the Hamiltonian. We note that configuration-interaction methodologies peculiar to ultracold systems with contact interactions have been developed \cite{Koscik2018,Koscik2020}, yet we will compare to a more general FCI method as the latter is more often utilised and is more general in scope.

FCI calculations are infeasible for large systems, since an upper bound to the scaling of the computational complexity of the FCI problem is $(K!/(N/2)!\,(K-N/2)!)^2$ \cite{Olsen1990}. For such systems where the FCI result is not available, we compare to the energy as found from coupled-cluster (CC) method \cite{Coester1958,Cizek1966,Cizek1971,Paldus1972,Bartlett1981,Bartlett1989,Bartlett2007}. This method utilises the HF ground-state and accounts for correlation through the use of a cluster operator, which acts on the reference wavefunction and produces linear combinations of excitations \cite{Bartlett2007}. The inclusion of single and double excitations is called the CCSD method, and if we further include approximate contributions from triple excitations which are found using many-body perturbation theory, we have the CCSD(T) method, which can be regarded as the gold standard in terms of efficiency and accuracy for small to medium size systems. Note that the CC method has been used for the one-dimensional ultracold few-fermion system extensively in the work by Grining \textit{et al.} \cite{Grining2015,Grining2015a}. In this work, the FCI, CCSD, and CCSD(T) results were found by using the PySCF program, a Python-based quantum chemistry package \cite{PYSCF} where we customised the Hamiltonian for our purposes. 

Because of the highly-degenerate nature of the larger systems considered in this work the ground-state energy itself does not serve as a conclusive metric of the accuracy of the RDM energy at larger interaction strengths nor for larger systems \cite{Volosniev2014,Deuretzbacher2014,minguzzi2022strongly}. As such, a comparison to known results in the field in such a case is pertinent. In the presence of infinitely strong repulsive interactions, i.e., when $g_{\mathsf{1D}}\to+\infty$, the $N$-fermion wavefunction can be derived exactly \cite{Girardeau1960,Guan2013,Volosniev2014,Deuretzbacher2014,minguzzi2022strongly}. This is because the infinitely strong repulsive interaction between two fermions can be cast as a manifestation of the Pauli exclusion principle, except applying to fermions without the same quantum numbers. Then, any wavefunction of the form $\Psi(r_1,r_2,\dots,r_N)$, where $r_i$ denotes the position of the $i$th fermion, necessarily satisfies $\Psi=0$ if $r_i=r_j$ for any $i,j\in\{1,2,\dots,N\}$. Note that each coordinate $r_i$ corresponds to a fermion which is either spin-up or spin-down, and this must be considered when calculating the explicit wavefunction. However, for the sake of simplicity, we do not incorporate explicit references to spin in the notation. Then, the wavefunction is proportional to the anti-symmetric Slater determinant,
\begin{align}
\Psi_{\mathsf{A}}(r_1,r_2,\dots,r_N)=\sum_{\sigma\in S_N}\mathsf{sgn}(\sigma)\prod_{i=1}^N\varphi_{i-1}(r_{\sigma_i}),
\end{align}
where $S_N$ is the symmetric group defined on a set of size $N$ and hence $\sigma$ is any permutation of the elements $\{1,2,\dots,N\}$, $\mathsf{sgn}(\sigma)$ is the parity of the permutation ($+1$ if it takes an even number of pairwise swaps to return the numbers into the original order $1,2,\dots,N$ and $-1$ otherwise) and $\varphi_i(r_{\sigma_i})$ is the quantum harmonic oscillator eigenfunction defined in Eq.~\eqref{eq:basis}. The general wavefunction for an $N$-fermion system is then expressed as \cite{Volosniev2014}

\begin{multline}
\Psi(r_1,r_2,\dots,r_N)=\sum_{k=1}^{N!}\,a_k\,\theta(r_{p_k(1)},r_{p_k(2)},\dots,r_{p_k(N)}) \\ \times\Psi_{\mathsf{A}}(r_1,r_2,\dots,r_N),\label{eq:analytic_wfn}
\end{multline}
where we sum over all $N!$ permutations of the coordinates $\{r_1,r_2,\dots,r_N\}$, and $\theta(r_1,r_2,\dots,r_N)=1$ if $r_1<r_2<\cdots<r_N$ and 0 otherwise and $a_k$ are simple coefficients. This solution was, for example, given explicitly for the $N=2+2$ case in Volosniev \textit{et al} \cite{Volosniev2014}. Note here we have the simple configuration in which the net spin is zero.

From this exact wavefunction, we can analytically calculate the density and pair-correlation function for $N$-fermion systems in the presence of infinitely strong repulsive interactions, thus permitting us to precisely assess the accuracy of the RDM methodology in capturing observable quantities in addition to the ground-state energy in the strongly-interacting regime. These results will be discussed in the next section, with the explicit densities and pair correlations, given in Fig.~\ref{fig:densities} and Fig.~\ref{fig:correlations}.

\section{Results}\label{sec:result}

The RDM methodology equips us with the ground-state energy as well as the linear expansion coefficients of the 2-RDM in the given basis. As such, not only can we compare the ground-state energy to other, more established methodologies in the field, but we can also construct other physical observables such as the single-particle density and the pair-correlation function. In this section, we first plot the ground-state energies of various ultracold, one-dimensional systems and compare to other methodologies, and then we investigate the single-particle densities and pair-correlation functions for different interaction strengths in both the attractive and repulsive regimes. Moreover, we extract the eigenvalues of the 2-RDM which permits us to analyse any pairing of fermions of opposite spin in the system. Note that in all plots presented the rank, $K$, refers to the spin-orbital basis rank.

\subsection{Ground-State Energies}

Here, we adopt the notation of $N=N_{\uparrow}+N_{\downarrow}$; thus, for example, the $N=1+1$ system is the two fermion system with one spin-up and one spin-down fermion. 

\begin{figure}[h!]
	\centering
	\includegraphics[width=0.8\linewidth]{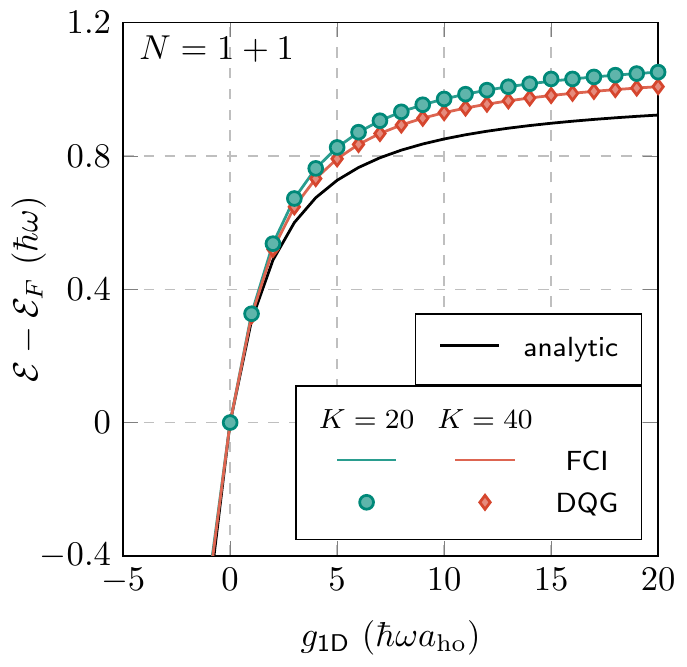}
	\caption{The ground-state energy (relative to the energy of the non-interacting system, $\mathcal{E}_F$) for the $N=1+1$ particle system. The RDM method with the \textsf{D}, \textsf{Q}, and \textsf{G} conditions enforced are given by the circles and diamonds, while the dashed lines indicate the exact solution. The analytic solution given by Busch \textit{et al.} \cite{Busch1998} is given by the black line. In green (circles) and red (diamonds) are the results for $K=20$ and $K=40$, respectively.}
	\label{fig:groundstate1}
\end{figure}

In Fig.~\ref{fig:groundstate1}, we consider the most simple system that demonstrates the efficacy of the RDM method where we plot the ground-state energy, relative to the energy of the non-interacting state $\mathcal{E}_F$, of the $N=1+1$ system as a function of the interaction strength $g_{\mathsf{1D}}$. A direct comparison between the FCI energy (solid lines) and the RDM energy with the \textsf{D}, \textsf{Q}, and \textsf{G} conditions enforced (circles and diamonds) is made, and it is seen that the agreement is excellent, with the magnitude of difference between the FCI and RDM energies being of the order of $10^{-3}\,\hbar\omega$ for all values of $g_{\mathsf{1D}}$ given, as we can see in Tab.~\ref{table:numericals1} in Appendix ~\ref{section:numerical_results}. This is expected, as the \textsf{D}, \textsf{Q}, and \textsf{G} conditions on the 2-RDM for an $N=2$ system constitute a complete set, and the resulting energy is exact, relative to the finite basis set, with the 2-RDM completely capturing all correlation in the system. 

For reference, the analytic solution for the $N=1+1$ system, as derived by Busch \textit{et al} \cite{Busch1998}, is shown by the black solid line. Naturally, we expect a difference in the predicted ground-state energy of the analytic and variational results for two reasons: firstly, we are approximating a function, i.e., the exact wavefunction, which is defined on an infinite-dimensional Hilbert space as a linear expansion of a finite number of basis functions and, secondly, the exact wavefunction possesses pathological behaviour such as cusps which are not aptly modelled the continuous basis functions. Here, a comparison to the analytic results is not the objective, but rather a comparison to the exact result in the finite basis which is the result given by the FCI calculation and the match, in this case, between the FCI and RDM energy is excellent.
\begin{figure*}
	\includegraphics[width=0.4\linewidth]{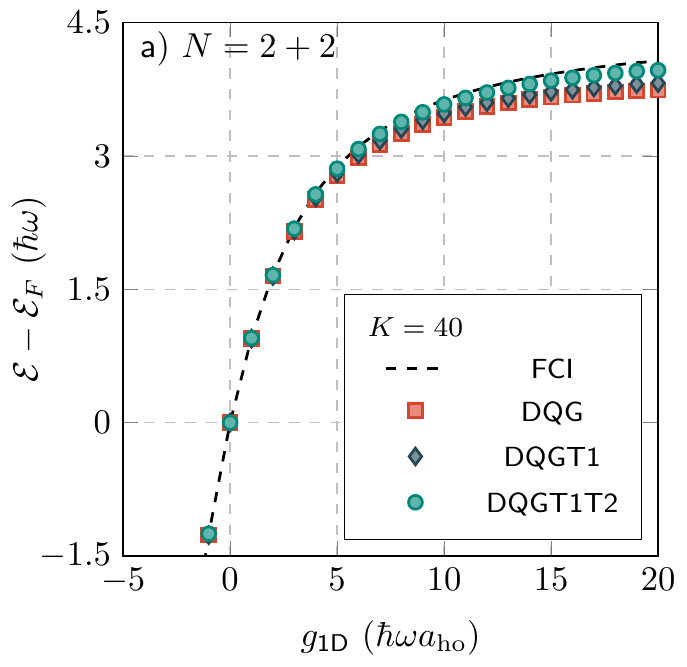}\includegraphics[width=0.4\linewidth]{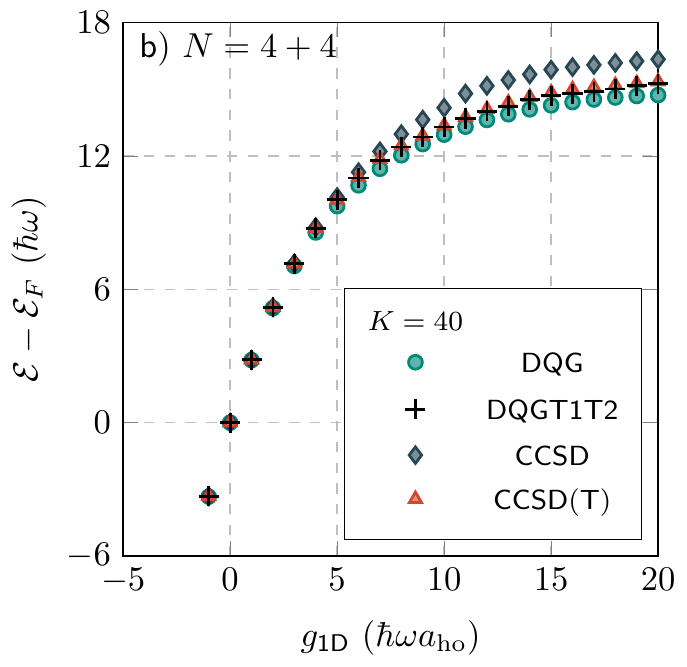}\\
	\includegraphics[width=0.4\linewidth]{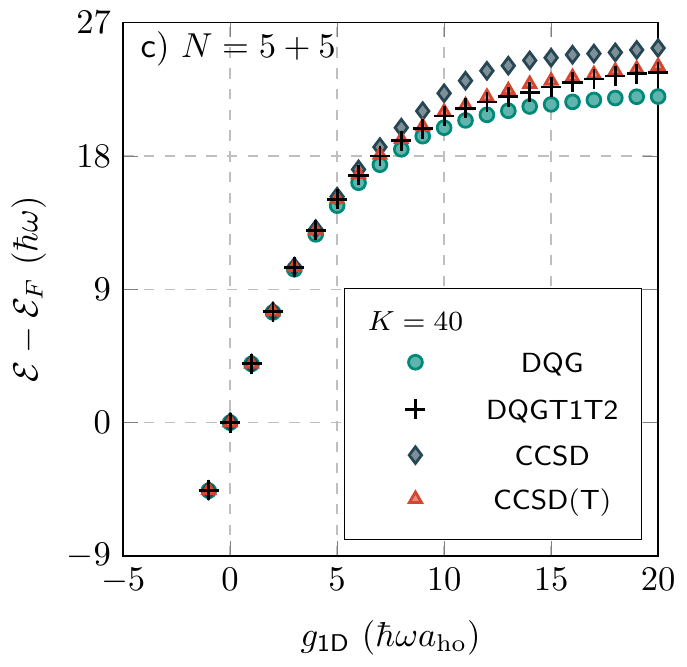}\includegraphics[width=0.4\linewidth]{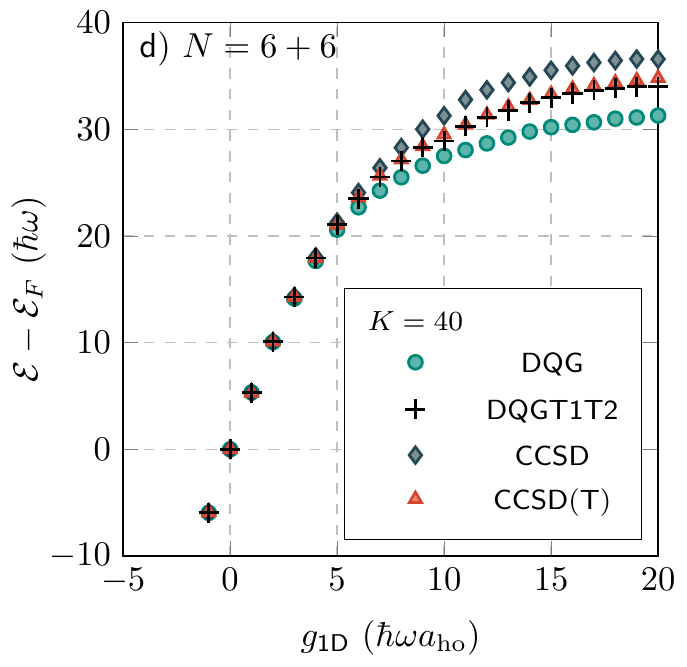}
	\caption{The ground-state energies (relative, in each case, to the corresponding energy of the non-interacting system, $\mathcal{E}_F$) of the a) $N=2+2$, b) $N=4+4$, c) $N=5+5$, and d) $N=6+6$ systems. We compare the energies found through the RDM method with the \textsf{D}, \textsf{Q}, and \textsf{G} conditions simultaneously enforced (circles) and then demonstrate the shift in energy found when the additional \textsf{T1} and \textsf{T2} conditions are enforced (black `$+$' signs). In the $N=2+2$ system (panel a)) we utilise the FCI energy as a benchmark to compare the accuracy (and also include the energy without the \textsf{T2} condition enforced) and in the remaining panels we used the CCSD and CCSD(T) energies as a comparison. Examples of the numerical differences are given in Tab.~\ref{table:numericals1} and Tab.~\ref{table:numericals2} in Appendix~\ref{section:numerical_results}.}
	\label{fig:groundstates2}
\end{figure*}

In Fig.~\ref{fig:groundstates2} we plot the ground-state energy, relative to the ground-state energy of the non-interacting system, $\mathcal{E}_F$, for the $N=2+2$, $4+4$, $5+5$, and $6+6$ systems. We make the comparison between the energy found by the RDM method (with the \textsf{D}, \textsf{Q}, and \textsf{G} initially enforced and then the \textsf{T1} and \textsf{T2} enforced also) and the FCI energy in the $N=2+2$ case (panel a)). We notice that the RDM method captures the energy accurately well into the strongly-interacting regime. Moreover, for the remaining systems, we compared the RDM energies to the CCSD and CCSD(T) energies and also see that the match between the CCSD(T) and RDM energy with all constraints up to the \textsf{T2} condition is excellent. Such an excellent agreement has also been demonstrated for various atoms and molecules \cite{Zhao2004}. Explicit numerical data detailing this agreement is given in Tab.~\ref{table:numericals1} and Tab.~\ref{table:numericals2} in Appendix~\ref{section:numerical_results}.

The systems exhibited here, particularly the $N=5+5$ and $N=6+6$ systems, exhibit strong correlation due to the number of interacting fermions and this is augmented as the interaction strength becomes very large. However, the RDM methodology accurately captures this correlation energy, as compared to the gold-standard CCSD(T) method. This utility of the RDM method, in addition to accurately capturing the correlation in these relatively large systems, is also manifest in the fact that the fermion number, $N$, enters the variational procedure merely as a parameter and has no bearing on the computational complexity, which only scales with rank. Thus, we see that the RDM scales well in its ability to capture strong correlation in larger systems with no additional computational resources required (if the calculations being compared are completed with the same spin-orbital basis rank, $K$). 

\begin{figure}[h]
	\centering
	\includegraphics[width=0.85\linewidth]{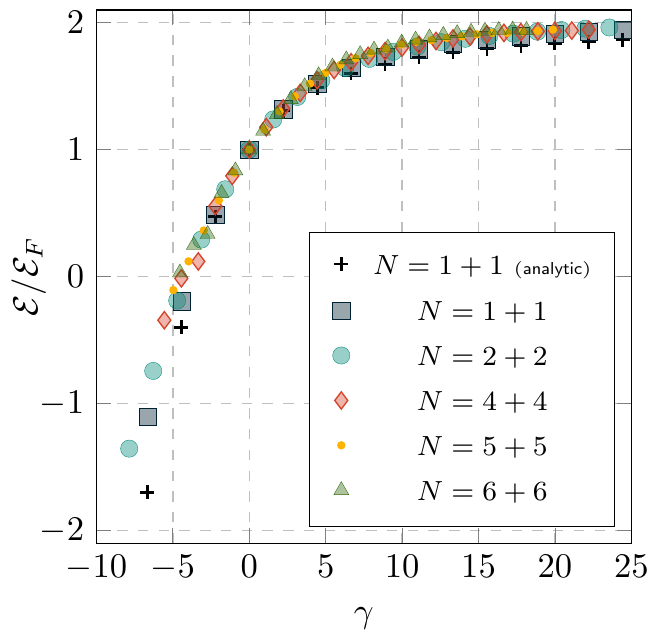}
	\caption{Rescaled energy, $\mathcal{E}/\mathcal{E}_F$, plotted against the rescaled interaction strength, $\gamma$, for the fermionic systems. The black `$+$' signs indicate the exact solution derived by Busch \textit{et al} \cite{Busch1998}. }
	\label{fig:comparison}
\end{figure}

A comparison between the ground-state results in each of these systems can be made by plotting the energy, $\mathcal{E}/\mathcal{E}_F$, where $\mathcal{E}$ is the ground-state energy and $\mathcal{E}_F$ is the energy of the non-interacting system, against a rescaled interaction strength, defined by $\gamma\coloneqq \pi g_{\mathsf{1D}}/\sqrt{N}$. Such an approach was given by Grining \textit{et al} \cite{Grining2015a} wherein it was shown that the energy difference between the $N=1+1$ system and the rescaled energy in the thermodynamic limit of an infinite number of fermions was extremely close. In Fig.~\ref{fig:comparison} we plot the rescaled energy against the rescaled interaction strength. We notice the significant overlap between the energies for each of the systems considered, which is consistent with the results given by Grining \textit{et al} \cite{Grining2015a}. Such an agreement extends well into repulsive and attractive interaction regimes.

Since CCSD methodologies have been shown to accurately reproduce the correlation energies in these systems \cite{Grining2015,Grining2015a} and since the RDM method matches significantly well with the CCSD results, we can couple the ground-state energies given in the panels in Fig.~\ref{fig:groundstates2} and the rescaled energies in Fig.~\ref{fig:comparison} to decisively state that the correlation in these strongly correlated systems is accurately captured by the \textsf{T1} and \textsf{T2} $N$-representability conditions in the RDM method.

\subsection{Densities, Pair-Correlations, and Eigenvalues}

In addition to the ground-state energy, we can also analyse any other physical observable as they all depend on the elements of the 1- and 2-RDMs. The density of the physical systems, which is defined as the diagonal elements of the 1-RDM,
\begin{align}
^{(1)}\rho(\bm x)\coloneqq\,^{(1)}D(\bm x;\bm x),
\end{align}
can be found from the elements of $^{(1)}D^i_j$ from Eq.~\eqref{eq:1rdm_basis}. The density profiles of these ultracold systems can be measured experimentally and are key observables as they define the physical extent of the interacting system and are a key indicator of the nature and strength of the interactions in the system.

\begin{figure}[h]
	\centering
	\includegraphics[width=0.85\linewidth]{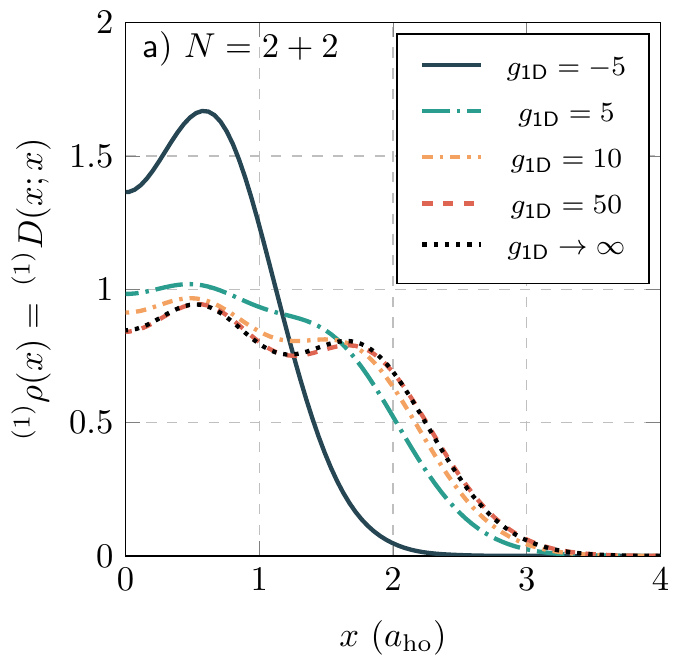}
	\includegraphics[width=0.85\linewidth]{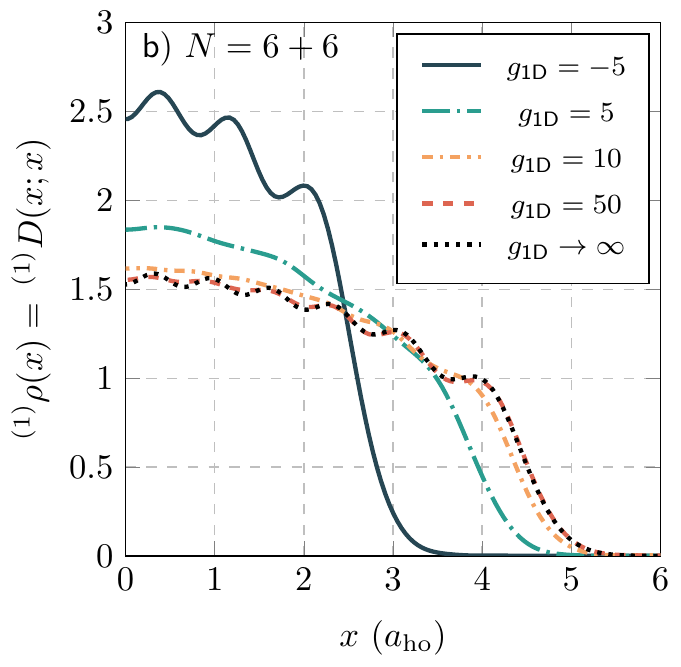}	
	\caption{The density, which is equal to the diagonal element of the 1-RDM, $\rho(x)=\,^{(1)}D(x;x)$, for the $N=2+2$ (panel a)) and the $N=6+6$ (panel b)) systems, plotted as a function of $x$ for four different interaction strengths, $g=-5, 5, 10$, and $50$. In all cases $K=40$ and the \textsf{D, Q, G, T1} and \textsf{T2} conditions are all applied. The density is normalised to the fermion number $N$. The density derived from the analytic wavefunction in the presence of infinitely strong interactions is given by the black dotted line in both cases.}
	\label{fig:densities}
\end{figure}

In Fig.~\ref{fig:densities} we plot the density profile of the $N=2+2$ (panel a)) and $N=6+6$ (panel b)) systems, respectively. In both, the spin-orbital basis rank is $K=40$ and the \textsf{D, Q, G, T1} and \textsf{T2} conditions are all applied. We find the density profiles for a range of different interaction strengths, from the attractive regime at $g_{\mathsf{1D}}=-5$ to the repulsive regime at $g_{\mathsf{1D}}=5,10$, and $50$. We also plot the analytic density in the limit that $g_{\mathsf{1D}}\to\infty$ which we can derive directly from the analytic wavefunction in Eq.~\eqref{eq:analytic_wfn} by integrating over the modulus squared of the wavefunction
\begin{multline}
^{(1)}\rho_{\mathsf{exact}}(r)\propto\int_{\mathbbm{R}^{N-1}}\text{d}r_2\cdots\text{d}r_N\,\Psi(r_1,\dots,R_N) \\ \times\Psi^*(r_1,\dots,R_N),
\end{multline}
where the proportionality accounts for the normalisation. The dotted line indicating the analytic density in Fig.~\ref{fig:densities} matches well with the density found by the RDM method at $g_{\mathsf{1D}}=50$, indicating that such a physical observable is aptly described by the RDM method for both the $N=2+2$ and $N=6+6$ systems, even in the relatively small basis rank of $K=40$.

The range of interaction strengths assessed in these density plots is sufficient to observe the impacts of attractive or repulsive interactions have on the system. We see a thickening of the tails as $g_{\mathsf{1D}}$ attains larger, positive values, which is to be expected in a system with strong repulsive interactions. This is contrasted with the density profile in the attractive interaction regime, with the width being smaller and the system attaining a high peak density closer to the centre of the trap. Such effects are clearly more pronounced in the $N=6+6$ system relative to the $N=2+2$ system. 

Another key physical observable is the pair-correlation function, which encodes all of the information about the pair-wise interactions the system exhibits. We define this generally as the diagonal element of the 2-RDM,
\begin{align}
^{(2)}\rho(\bm x_1,\bm x_2)\coloneqq\,^{(2)}D(\bm x_1,\bm x_2;\bm x_1,\bm x_2).
\end{align}
There is inherent ambiguity in the definition of $^{(2)}\rho$ in this case, as the 2-RDM $^{(2)}D(\bm x_1,\bm x_2;\bm x_1',\bm x_2')$ generally encodes information about the correlation between particles of the same species (which do not occur in this system due to Pauli exclusion and the interactions being point-like) as well as between different species. Since we are considering contact interactions, it is the latter components of the 2-RDM which we consider and thus we only sum over the appropriate indices in Eq.~\eqref{eq:2rdm_basis} to garner the pair-correlation function. As in the case of the densities, we can also calculate the analytic pair-correlation function in the limit $g_{\mathsf{1D}}\to\infty$ directly from the definition of the wavefunction in Eq.~\eqref{eq:analytic_wfn}.

\begin{figure}[h]
	\centering
	\includegraphics[width=0.85\linewidth]{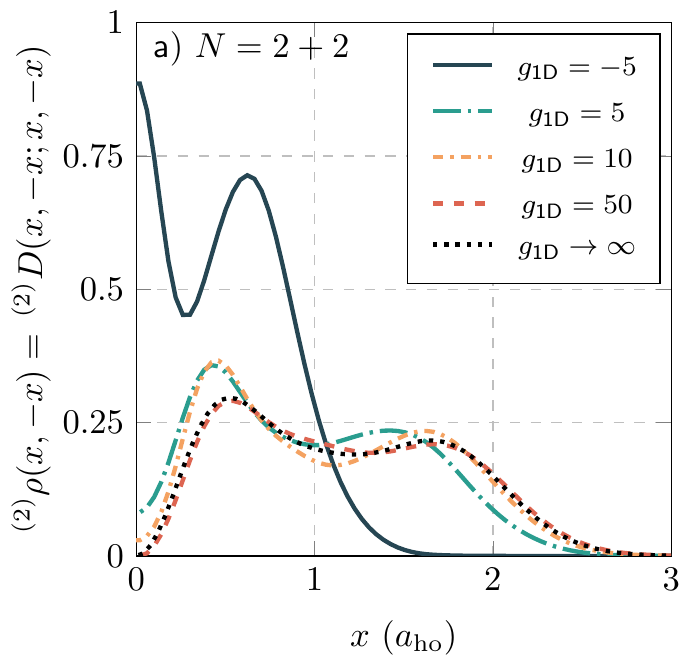}
	\includegraphics[width=0.85\linewidth]{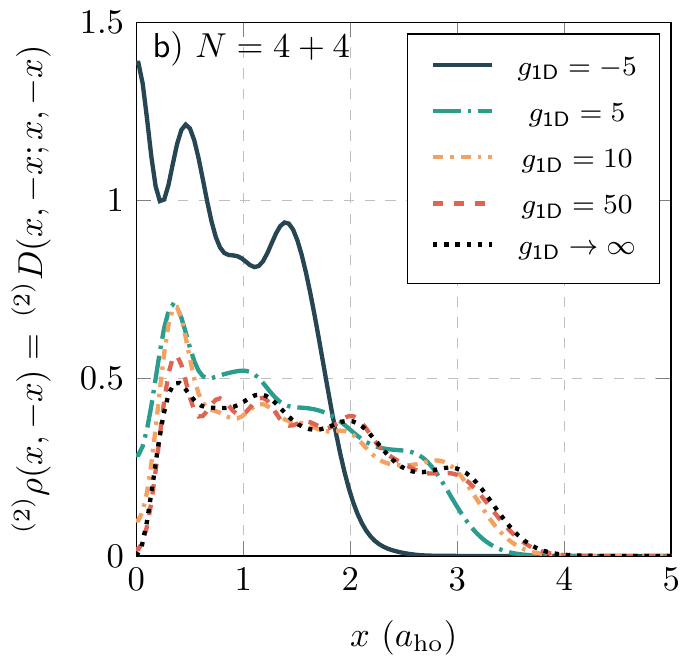}	
	\caption{The pair-correlation function, $^{(2)}\rho(x,-x)$, for the $N=2+2$ (panel a)) and $N=4+4$ (panel b)) systems, plotted as a function of $x$ for four different interaction strengths, $g=-5,5,10,$ and $50$. In all cases $r=40$ and the \textsf{D, Q, G, T1} and \textsf{T2} conditions are all applied.}
	\label{fig:correlations}
\end{figure}

In Fig.~\ref{fig:correlations} we plot the `anti-diagonal' component of the pair-correlation function, $^{(2)}\rho(x,-x)$, for the $N=2+2$ (panel a)) and $N=4+4$ (panel b)) systems, for a range of values of the contact interaction strength, from the attractive regime at $g=-5$ into the repulsive regime at $g=5,10$, and $g=50$. We also include the analytic pair-correlation in the limit that $g_{\mathsf{1D}}\to\infty$, which is indicated by the dotted black line in both cases. We notice the ability of the RDM methodology in capturing the pair-correlation in the strongly interacting limit by comparing the RDM pair-correlation at $g_{\mathsf{1D}}=50$ to the exact solution at $g_{\mathsf{1D}}\to\infty$.

In all cases, the spin-orbital basis rank is $K=40$ and the \textsf{D, Q, G, T1} and \textsf{T2} conditions are simultaneously enforced. The pair-correlation function, as defined as the diagonal element of the 2-RDM, is a probability distribution function for two particles (of opposite spin, in this case) occupying positions $r_1$ and $r_2$, and hence the anti-diagonal component thereof is readily interpreted as a probability distribution function for two opposite spin particles to be occupying positions equidistant from the centre of the trap. As we can see, the attractive interactions significantly increase the likelihood with which two particles will be found in mutual proximity (near the centre of the trap). 

A final characteristic we consider in this work is the spectral decomposition of the 2-RDM. The eigenvalues of the components of the 2-RDM corresponding to the interaction between particles of opposite spin gives us insight into the existence of pairing in the system. The archetypal example of pairing in fermionic system is the existence of Cooper pairs in low-temperature electronic systems \cite{Cooper1956}, which is the driving mechanism for the superconductivity as explained in the Bardeen-Cooper-Schreiffer (BCS) theory \cite{Bardeen1957,Bardeen1957a}. However, other pairing mechanisms in ultracold system have been studied both theoretically and experimentally \cite{Bohr1958,Migdal1959,Yang1971,Viverit2004,Juillet2004}. The ultracold systems analysed in this work are known to exhibit pairing \cite{Sowinski2015} and here we demonstrate that the RDM methodology, in addition to capturing the salient features of these systems as already discussed, accurately predicts the occurrence of such a phenomenon. 
 
\begin{figure}[h]
	\centering
	\includegraphics[width=0.85\linewidth]{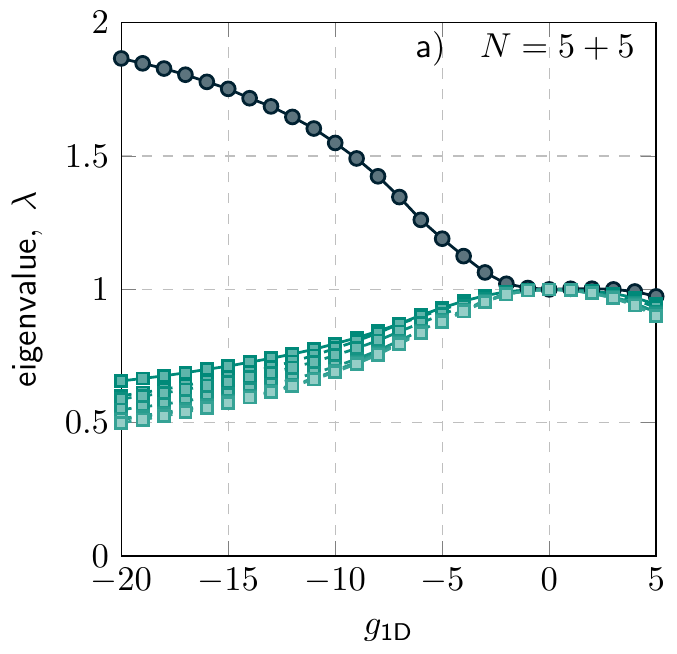}
	\includegraphics[width=0.85\linewidth]{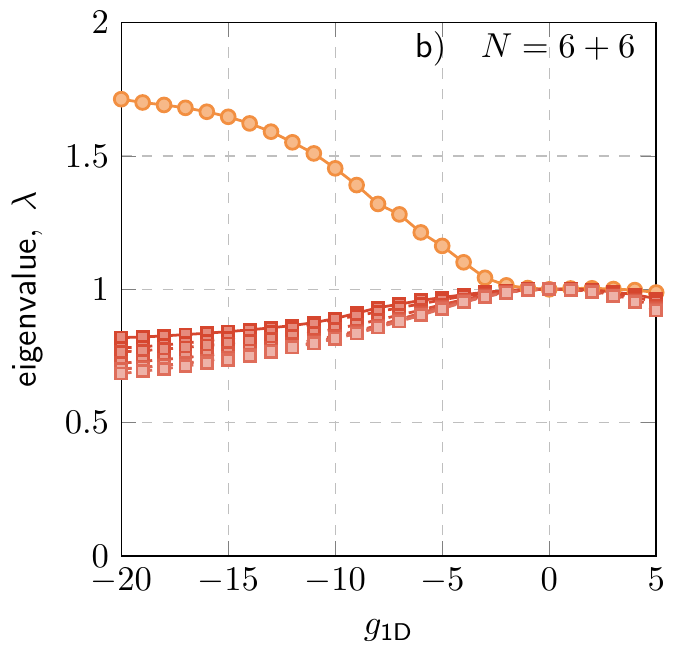}	
	\caption{The first eight eigenvalues of the 2-RDM, $^{(2)}D(x_1,x_2;x_1',x_2')$, where $x_1,x_1'$ are spin-up and $x_2,x_2'$ are spin-down, for the $N=5+5$ (panel \textsf{a)}) and $N=6+6$ (panel \textsf{b)}) systems. In both cases, $r=40$ and the \textsf{D}, \textsf{Q}, \textsf{G}, \textsf{T1}, and \textsf{T2} conditions are simultaneously enforced. The highest eigenvalues at each interaction strength is denoted by the circles, with the remaining seven indicated by the squares.}
	\label{fig:eigen_plot}
\end{figure}

In Fig.~\ref{fig:eigen_plot} we plot the first eight eigenvalues of the 2-RDM as a function of the interaction strength in the attractive regime for both the $N=5+5$ (Fig.~\ref{fig:eigen_plot} a)) and $N=6+6$ (Fig.~\ref{fig:eigen_plot} b)) systems. In both cases, the first-order eigenvalue for each interaction strength is indicated by the circles and the remaining seven are indicated by the squares. We note that a single eigenvalue dominates above the others, indicating a BCS-like pairing manifesting for two opposite-spin particles occupying the same spatial orbital \cite{Lydzba2020}. Such results are in agreement with previous results demonstrated in the literature \cite{Sowinski2015} and, moreover, are consistent with the behaviour of the pair-correlation function for attractive interactions as in Fig.~\ref{fig:correlations}, in which the likelihood of two opposite-spin particles being proximate to one another is significant. 

\subsection{A Note on Efficiency}

A key facet of the effectiveness of the RDM method is the relative lack of memory and time requirements when it comes to computational implementation compared to the FCI method, as the RDM method scales polynomially in the basis rank and does not scale with particle number at all (with the FCI method scaling exponentially in terms of the basis rank \textit{and} the particle number, at best \cite{Aquilante2016}). The computational complexity of the interior-point methods scaled approximately as $K^{16}$, where $K$ is the basis rank. Such a poor scaling meant that the method was confined to small atomic and molecular systems. However, faster algorithms, such as those used by SDPNAL$+$ result in a scaling of approximately $K^6$ to $K^8$. Such an improvement in the scaling of the complexity was also manifest in the boundary-point method developed by Mazziotti \cite{Mazziotti2011}, which demonstrated a ten- to twenty-fold increase in efficiency to other calculations at the time.

\begin{table}[htb]
\centering
\renewcommand\multirowsetup{\centering}
\caption{CPU time (in minutes) for various systems and spin-orbital basis ranks for the 2-RDM and FCI calculations. All 2-RDM calculations simultaneously enforced the \textsf{D, Q} and \textsf{G} conditions. We note that the 2-RDM calculations scale only as a function of the rank $K$, and hence only one calculation need be performed for each, whereas the FCI method scales exponentially with both basis size and rank, as seen for each rank as we increase the particle number. Calculations were performed on a laptop with Intel Core i7 2.60GHz processors (64GB RAM) using 6 cores. The accuracy of the sample 2-RDM calculations is indicated by the percentage of the correlation energy (where the correlation energy is defined by $E_{\mathsf{corr}}=E_{\mathsf{HF}}-E_{\mathsf{FCI}}$, i.e., the difference between the mean-field HF energy and the exact FCI energy). In all calculations, $g_{\mathsf{1D}}=10$. } 
\begin{tabular}{ccccc}

	\toprule\midrule
	\multirow{ 2}{*}{\textsf{rank}, $K$} & \multirow{ 2}{*}{\textsf{fermions}, $N$} & \multicolumn{2}{c}{\textsf{time (m)}} & \textsf{RDM}  \%$E_{\mathsf{corr}}$\\\cline{3-4}\\[-0.5em]
	& & \textsf{FCI} & \textsf{RDM} &   \\\midrule
	 & 2 & 0.001 & & 100.00\\
	20 & 4 & 0.006 & 1.298 & 101.05\\
	& 8 & 0.112 & & 117.32\\\midrule
	& 2 & 0.003 & & 100.00\\
	30 & 4 & 0.072 & 5.125 & 102.97\\
	& 8 & 11.702 & & 116.68\\\midrule
	& 2 & 0.042 & & 100.00\\
	40 & 4 & 0.790 & 17.3482 & 102.95\\
	& 8 & 152.674 & & 116.34 \\\midrule\bottomrule
\end{tabular}
\label{tab:tab1}
\end{table}

To demonstrate the efficiency, we found, in Tab.~\ref{tab:tab1}, the CPU time (in minutes) for various calculations, and the dependency of the CPU time on fermion number and basis rank for the 2-RDM and FCI calculations, as well as the percentage of the correlation energy captured by the 2-RDM method, with the FCI energy being defined as 100\% correlated. The 2-RDM method does not scale with fermion number, which is an enormous advantage over methods that require to calculation of the $N$-fermion wavefunction. We can see the large increase in CPU time for large rank and fermion number for the FCI calculations, with the relatively slow increase with rank of the 2-RDM calculations. All 2-RDM calculations simultaneously enforce the \textsf{D}, \textsf{Q}, and \textsf{G} conditions. 

Enforcing further constraints on the 2-RDM, such as the \textsf{T1} and \textsf{T2} constraints results in a much larger SDP and hence a larger CPU time. Generally speaking, the largest hindrance to the widespread adoption of this method was the poor scaling of the computational complexity with basis rank as one increased the amount of $N$-representability conditions enforced. However, recent effort has significantly reduced this scaling from approximately $K^9$ to $K^6$ \cite{Mazziotti2016,Mazziotti2020}.

\section{Concluding Remarks}

In this work, we outlined the theoretical development of the RDM method, as well as briefly demonstrating how to complete a simple implementation via a formulation as an SDP. Such a method has seen wide success in atomic and molecular applications, and here we extend the range of applicability by considering a simple, highly correlated few-fermion system. Specifically, we applied the RDM method with the \textsf{D, Q, G, T1,} and \textsf{T2} $N$-representability constraints to a trapped, one-dimensional system of fermions interacting via a contact interaction, while constraining the system to have an equal number of spin-up and spin-down fermions. The ground-state energy was found for $N=2,4,8,10$ and $12$ particle systems and was found to match well, for a range of interaction strengths in the positive and negative regimes, with the best theoretical results in the given harmonic oscillator, basis, provided by the FCI or exact energy for the $N=2$ and $N=4$ system, and by the CCSD and CCSD(T) method for the $N=8, 10,$ and $12$ system. 

Specifically, for the $N=1+1$ case, the ground-state energy given by the RDM method with the \textsf{DQG} conditions implemented agreed with the exact FCI energy for all interaction strengths, $-5\le g\le 20$, which is expected. In the $N=2+2, 4+4, 5+5$ and $N=6+6$ cases, it was demonstrated that while constraining the 2-RDM with the \textsf{D, Q, G, T1,} and \textsf{T2} conditions, strong correlation was accurately captured, with the ground-state energy being close to that given by FCI and CCSD(T) methods. This demonstrates the effectiveness of the RDM method when it comes to capturing the ground-state energies in strongly correlated systems, and further demonstrates that this method is an extremely promising alternative to typical methods based on the complete knowledge of the $N$-fermion wavefunction. Moreover, we demonstrated that the three-body $N$-representability constraints are sufficient in a variational procedure to minimise the 2-RDM, and higher-order constraints need not be considered. 

The density and components of the pair-correlation function we also considered for the $N=2+2$ and $N=6+6$, and $N=2+2$ and $N=4+4$ systems, respectively, with a basis rank of $K=40$ and simultaneous enforcement of the \textsf{D, Q, G, T1} and \textsf{T2} conditions. These quantities are calculated directly from the variationally determined components of the 1- and 2-RDMs. Such observables were found in the strongly interacting regime, where $g_{\mathsf{1D}}=50$, and were found to match well with the anayltic result for the observables in the limit of infinitely strong repulsive interactions. By means of Eq.~\eqref{eq:1rdm_basis} and Eq.~\eqref{eq:2rdm_basis} any physical observables can be readily expressed in terms of the components of $^{(1)}D$ and $^{(2)}D$. We also demonstrated that important features of these systems such as the phenomenon of BCS-like pairing was manifested in the spectral decomposition of the 2-RDM. As such, the key features of these systems are all found naturally by considering the variationally determined elements of the 1- and 2-RDMs.

The strong correlation in these one-dimensional systems, which occurs in the regime of $|g_{\mathsf{1D}}|\gg 1$, was accurately captured by the RDM method after implementing the three-body constraints known as the \textsf{T1} and \textsf{T2} conditions. This shows that, in addition to the RDM method accurately capturing strong electron correlation in atomic and molecular systems, the RDM method accurately accounts for strong correlation in ultracold few-fermion systems where $g_{\mathsf{1D}}$ can be made arbitrarily high. An important point to note, with regard to this method, is that the introduction of higher-order constraints provides a mechanism for understanding the limitations of lower order calculations. Additionally, for a given rank, utilisation of the RDM method in parallel with conventional wavefunction techniques can allow one to evaluate an upper and lower bound for the ground-state energy of the system.

The choice of basis, namely the harmonic oscillator basis, naturally hinders the attempt to find extremely accurate ground-state energies relative to the known, analytic case for the $N=1+1$ system, and other approach for larger $N$. However, what has been clearly demonstrated is that the RDM method gives accurate results relative to the FCI and CCSD(T) methods for relatively large systems and strong interaction strengths in the same basis.

The accuracy of the RDM method for these simple systems opens the door to applying the same method for more complex interacting systems. An extension to the analogous system studied here in two and three dimensions is a natural case for further investigation. Further extensions using this methodology could include spin imbalanced systems, where $N_{\uparrow}\neq N_{\downarrow}$, as well as mass-imbalanced systems. Minor imbalances in these systems can be exaggerated to permit the study of single impurities in otherwise homogeneous systems \cite{Cetina2016,Parish2016,Kerin2020}. Alternate interactions, such as soft-core or dipolar interactions can be analysed with this method also. Such changes amount to mere substitutions in the Hamiltonian matrix elements, and modifications of the particle number and net spin, which are simple parameters in the calculation and requires no modification of the methodology as a whole nor any increase in computational complexity. This demonstrates the universality of the RDM approach to many-body quantum systems with pairwise interactions and exemplifies why it is a promising method in the field. 

The 2-RDM approach adopted here achieves the goals of $N$-representability and universality that have driven the development of DFT, which is the 1-RDM method for fermion systems. The exponential complexity of $N$-fermion wavefunction-based approaches has been circumvented in favour of a method with polynomial scaling characteristics. The systematic order-by-order application of $N$-representability constraints leads to a computational scheme for 2-RDM that can be made as accurate as required; it is a ``gold standard'' comparable to full configuration interaction (FCI). Nevertheless, the computational cost of 2-RDM exhibits an intrinsic computational complexity far greater than that of DFT, and comparable with coupled-pair methods, such as the CC approximation. The 2-RDM method is, however, more than just a convenient computational tool. It provides direct insights into the relationship between the energy of a fermionic system and its 2-RDM through the simple process of integration over the coordinates of a single fermion. The physical principles that underpin the properties of $N$-fermion wavefunctions that are used to derive the Hohenberg-Kohn theorems of DFT apply with equal validity to an $N$-representable 2-RDM. This suggests the potential for a parallel development of 1-RDM methods based on the availability of 2-RDM approaches for complex fermion systems.

\begin{acknowledgments}
M. J. K. is  supported  by  an  Australian Government Research Training Program Scholarship and by the University of Melbourne. The authors thank S. B. Prasad for helpful feedback during the drafting of this paper.
\end{acknowledgments}

\appendix

\section{Numerical Results}\label{section:numerical_results}

In this appendix, we include the relevant numerical results that pertain to the ground-state energies as a function of the interaction strength as given in Fig.~\ref{fig:groundstate1} and Fig.~\ref{fig:groundstates2}. In Tab.~\ref{table:numericals1} and Tab.~\ref{table:numericals2} we give the ground-state energy captured by the RDM method as a percentage of the energy captured by the FCI (for the $N=1+1$ and $N=2+2$ systems) and the CCSD and CCSD(T) methods ($N=4+4$, $5+5$, and $6+6$) for the interactions of strength $g_{\mathsf{1D}}=-5$, $10$, and $20$.

The ground-state energy found by the RDM method is lower, in all cases, compared to the FCI or CCSD/CCSD(T) energies. This is indicative of the fact that if one adds successive $N$-representability constraints to the variational determination of the 2-RDM we approach the ground-state energy from \textit{below}. Thus, lower values of the percentage of the energy captured by the RDM method (when compared to the FCI or CCSD/CCSD(T) which we take as a benchmark) indicate that further, higher-order constraints are needed. It must be noted, however, that the numerical results further demonstrate the ability of the RDM method to capture the strong correlation, even in the relatively large $N=5+5$ and $N=6+6$ systems.

\begin{table}
	\centering
	\begin{tabular}{ccc|ccc}\toprule\midrule
		& & \multicolumn{1}{c}{} & \multicolumn{3}{c}{\textsf{energy (as \% of $E_{\mathsf{FCI}}$)}}\\\cline{4-6}
		$g_{\mathsf{1D}}$ & $N$ &\multicolumn{1}{c}{} $E_{\mathsf{FCI}}$ & $E_{\mathsf{DQG}}$ & $E_{\mathsf{DQGT1}}$ & $E_{\mathsf{DQGT1T2}}$\\\midrule
		\multirow{ 2}{*}{$-5$} & $1+1$ & -3.40734 & 99.9988 &       - & -\\
		                       & $2+2$ & -5.47014 & 97.7234 &	 98.9954 & 99.0859\\\midrule
		\multirow{ 2}{*}{$10$} & $1+1$ & 1.929660 & 99.9999 &       - & -\\
		                       & $2+2$ & 7.630125 & 97.4299 & 98.0300 & 99.3447\\\midrule
		\multirow{ 2}{*}{$20$} & $1+1$ & 2.007683	 & 99.9971 &       - & -\\
		                       & $2+2$ & 8.065181 & 95.9892 & 96.9682 & 98.7601\\\midrule\bottomrule              
	\end{tabular}
	\caption{Comparison between the energy found from the FCI calculation, $E_{\mathsf{FCI}}$, and the energy found by the RDM method, with the RDM energy given as a percentage of $E_{\mathsf{FCI}}$. Here, we consider the $N=1+1$ and $N=2+2$ systems for $g_{\mathsf{1D}}=-5, 10$, and $20$. }
	\label{table:numericals1}
\end{table}

\begin{table}
	\begin{tabular}{ccc|ccc}\toprule\midrule
		& & \multicolumn{1}{c}{} & \multicolumn{3}{c}{\textsf{energy (as \% of $E_{\mathsf{CCSD(T)}}$)}}\\\cline{4-6}
		$g_{\mathsf{1D}}$ & $N$ &\multicolumn{1}{c}{} $E_{\mathsf{CCSD(T)}}$ & $E_{\mathsf{DQG}}$ & $E_{\mathsf{DQGT1}}$ & $E_{\mathsf{DQGT1T2}}$\\\midrule
	          & $4+4$ & -5.61208 & 95.0938 & 96.5067 & 98.4608\\
		 $-5$ & $5+5$ & -2.65624 & 94.4669 & 96.9961 & 98.1003\\
		      & $6+6$ &  0.43422 & 93.7429 & 94.2232 & 97.6343\\\midrule
	          & $4+4$ & 29.36926 & 97.8916 & 98.4081 & 99.4324\\
		 $10$ & $5+5$ & 45.97926 & 97.6692 & 97.8424 & 99.3766\\
		      & $6+6$ & 65.50691 & 96.9456 & 97.3655 & 99.0678\\\midrule
	          & $4+4$ & 31.35000 & 96.0203 & 96.9120 & 99.2026\\
		 $20$ & $5+5$ & 49.06669 & 95.8112 & 96.4841 & 99.1217\\
		      & $6+6$ & 70.80834 &95.0337 & 96.9302 & 98.7001\\\midrule\bottomrule   
	\end{tabular}
	\caption{Comparison between the energy found from the CCSD(T) calculation, $E_{\mathsf{CCSD(T)}}$, and the energy found by the RDM method, with the RDM energy given as a percentage of $E_{\mathsf{CCSD(T)}}$. Here, we consider the $N=1+1$ and $N=2+2$ systems for $g_{\mathsf{1D}}=-5, 10$, and $20$.}
	\label{table:numericals2}
\end{table}

\bibliography{rdm_ultracold_few_fermion_systems_onedim.bib}

\end{document}